\documentclass[a4paper,11pt]{article}
\pdfoutput=1 

\usepackage{jheppub} 

\usepackage[T1]{fontenc} 

\usepackage{xcolor}
\usepackage{datetime}
\DeclareUnicodeCharacter{2212}{-}

\title{\boldmath Composite Topological Structures in $\mathbf{SO(10)}$}

\author[a,1]{George Lazarides,\note{\href{mailto:glazarid@gen.auth.gr}{glazarid@gen.auth.gr}}}
\author[b,2]{Qaisar Shafi\note{\href{mailto:qshafi@udel.edu}{qshafi@udel.edu}}}
\author[b,3]{and Amit Tiwari\note{\href{mailto:amitiit@udel.edu}{amitiit@udel.edu}}}

\affiliation[a]{School of Electrical and Computer Engineering, Faculty of Engineering,
Aristotle University of Thessaloniki, Thessaloniki 54124, Greece}
\affiliation[b]{Bartol Research Institute, Department of Physics and Astronomy,
University of Delaware, Newark, DE 19716, USA}

\abstract{We explore a variety of composite topological structures that arise from the spontaneous breaking of $SO(10)$ to $SU(3)_c \times U(1)_{em}$ via one of its maximal subgroups $SU(5) \times U(1)_\chi$, $SU(4)_c \times SU(2)_L \times SU(2)_R$, and $SU(5) \times U(1)_X$ (also known as flipped $SU(5)$). They include i) a network of $\Z$ strings which develop monopoles and turn into necklaces with the structure of $\Z_2$ strings, ii) dumbbells connecting two different types of monopoles, or monopoles and antimonpoles, iii) starfish-like configurations, iv) polypole configurations, and v) walls bounded by a necklace. We display these structures both before and after the electroweak breaking. The appearance of these composite structures in the early universe and their astrophysical implications including gravitational wave emission would depend on the symmetry breaking patterns and scales, and the nature of the associated phase transitions.}

\begin{document} 
\maketitle
\flushbottom

\section{Introduction}
\label{sec:intro}
In the late seventies and early eighties there arose an important challenge which was to search for topologically stable strings
in unified theories \cite{Pati:1974yy,Georgi:1974sy,Georgi:1974my,Fritzsch:1974nn}, without invoking an additional ad hoc $U(1)$ symmetry. This was settled with
the discovery \cite{Kibble:1982ae} of such strings in $SO(10)$, more precisely Spin(10), provided its spontaneous symmetry breaking to the Standard Model (SM) gauge group $G_{SM}=SU(3)_c\times SU(2)_L\times U(1)_Y$ and subsequently to $SU(3)_c \times U(1)_{em}$ is implemented with scalar fields in the tensor representations. [For simplicity of notation we use throughout a less precise notation for the $SO(10)$ subgroups. For instance, the SM subgroup $G_{SM}$ is written without the factor $\Z_3 \times \Z_2$ in the denominator, where $\Z_3$ and $\Z_2$ denote the centers of $SU(3)_c$ and $SU(2)_L$ respectively.] The breaking of $SO(10)$ with tensor representations, independent of the details on how the symmetry breaking is implemented, leaves unbroken the $\Z_2$ subgroup of $\Z_4$, the center of $SO(10)$ \cite{Kibble:1982ae}.

Around the same time, it was demonstrated that extended topological structures such as domain walls bounded by strings \cite{Kibble:1982dd} can also appear in theories such as $SO(10)$. The original example of this structure \cite{Kibble:1982dd} was based on the breaking of $SO(10)$ to $SU(4)_c\times SU(2)_L \times SU(2)_R$ with a 54-plet, which also leaves unbroken a discrete symmetry known as `$C$-parity'. As shown in Ref.~\cite{Lazarides:1985my}, $CQC^{-1}=-Q$, where $Q$ is the electric charge generator in $SO(10)$. The subsequent breaking of $C$-parity leads to the appearance of walls bounded by strings. It was shown in Ref.~\cite{Lazarides:1985my} that this domain wall is superconducting. Remarkably, walls bounded by strings have recently been found in superfluid helium $^{3}$He-B \cite{Makinen:2018ltj}.

The experimental discovery of gravitational waves predicted in Einstein’s general theory of relativity has triggered extensive investigations in the stochastic gravitational radiation emitted by cosmic strings \cite{Buchmuller:2019gfy,Buchmuller:2020lbh,Sousa:2020sxs,Blanco-Pillado:2021ygr,Lazarides:2021uxv,Buchmuller:2021mbb,Chakrabortty:2020otp,King:2020hyd,King:2021gmj,Lazarides:2022jgr,Afzal:2022vjx}, domain walls \cite{Borboruah:2022eex,Borah:2022wdy,Banerjee:2023hcx}, and composite structures such as walls bounded by strings \cite{Everett:1982nm,Dunsky:2021tih}. Therefore it is timely to explore in greater depth the variety of composite topological structures that can arise in realistic Grand Unified Theory (GUT) models based on $SO(10)$ and $E_6$ (for an earlier discussion, see Refs.~\cite{Jeannerot:2003qv,Lazarides:2019xai}). As an interesting example, in the breaking of $SO(10)$ to $G_{SM}$ via $SU(4)_c \times SU(2)_L \times SU(2)_R$, in addition to cosmic strings and an observable number density of intermediate mass monopoles \cite{Lazarides:1984pq,Chakrabortty:2020otp,Senoguz:2015lba,Maji:2022jzu}, we also find necklace configurations consisting of two distinct types of monopoles residing on a string \cite{Lazarides:2019xai}. It is worth noting that this type of composite topological structure (necklace) can also appear in superfluid $^{3}$He \cite{Volovik:2019goo,Volovik:2020zqc,Makinen:2022pqn}. [For earlier work on monopoles on strings, see Refs.~\cite{Vilenkin:1982hm,Hindmarsh:1985xc,Aryal:1987sn}, and for a review, Ref.~\cite{Kibble:2015twa} and references therein].

In this paper, following an earlier work \cite{Lazarides:2019xai} where we mainly focused on $SO(10)$ breaking via $SU(4)_c \times SU(2)_L \times SU(2)_R$, we show the presence of a variety of composite topological structures that may arise from the spontaneous breaking of $SO(10)$ to $SU(3)_c \times U(1)_{em}$ via its maximal subgroups $SU(5) \times U(1)_{\chi}$, $SU(4)_c \times SU(2)_L \times SU(2)_R$ \cite{Pati:1974yy}, and $SU(5) \times U(1)_X$ \cite{DeRujula:1980qc}, also known as flipped $SU(5)$ \cite{Barr:1981qv}. (Although there are many breaking chains of $SO(10)$, we concentrate on these three models since they are phenomenologically viable and lead to many new extended structures.) We display the presence of a variety of composite topological structures that include i) a network of $\Z$ strings which develop monopoles and turn into necklaces with the structure of $\Z_2$ strings, ii) dumbbells connecting two different types of monopoles, or monopoles and antimonpoles, iii) starfish-like configurations, iv) polypole configurations, and v) walls bounded by a necklace. We also display the form of these structures both before and after the electroweak (EW) breaking.

A careful exploration of the gravitational wave spectrum generated by these topological structures may shed light not only on the underlying GUT, $SO(10)$ in our case, but also on the symmetry breaking patterns as well as the nature of the scalar fields responsible for the breaking (for various $SO(10)$ symmetry breaking patterns see Refs.~\cite{Chakrabortty:2009xm,Chakrabortty:2017mgi,Chakrabortty:2019fov,King:2020hyd,King:2021gmj,Holman:1982tb,Ohlsson:2020rjc}). In particular, as previously mentioned, if the breaking is carried out with scalar field vacuum expectation values (VEVs) in the tensor representations, an unbroken $\Z_2$ symmetry survives which plays the role of ‘matter’ parity in a supersymmetric setting and which ensures the stability of the lightest supersymmetric particle (LSP) dark matter. In a non-supersymmetric framework, it can be used to stabilize a desired dark matter candidate. For an early example of this see Ref.~\cite{Holman:1982tb}, and more recent examples can be found in Refs.~\cite{Kadastik:2009dj,Mambrini:2015vna,Boucenna:2015sdg,Ferrari:2018rey,Lazarides:2020frf,Okada:2022yvq,Lazarides:2022ezc}. 

The spontaneous breaking of $\Z_2$ symmetry yields domain walls bounded by a necklace, as we show in later sections. It would be exciting if such objects can be realized in superfluid helium or some other condensed matter system.

In the following discussion, we consider first the breaking of $SO(10)$ via
$SU(5) \times U(1)_{\chi}$ and show how the composite topological structures mentioned above appear. The impact on these structures following the EW breaking is also considered. [For an early discussion of $SO(10)$ vortices and the EW phase transition, see Ref.~\cite{Stern:1985bg}]. This discussion is followed by considerations of the other two symmetry breaking chains of $SO(10)$, namely breaking via $SU(4)_c \times SU(2)_L \times SU(2)_R$ and flipped $SU(5)$.

\section{$\mathbf{SO(10)}$ breaking via $\mathbf{SU(5)\times U(1)_{\chi}}$}
\label{sec:SU5chi}
Consider the following breaking of $SO(10)$ to $SU(3)_c \times U(1)_{em}$:

\begin{equation}
\begin{split}
    SO(10) &\longrightarrow SU(5) \times U(1)_\chi\\
    &\longrightarrow
SU(3)_c \times SU(2)_L \times U(1)_Y \times U(1)_\chi\\
&\longrightarrow SU(3)_c \times SU(2)_L \times U(1)_Y\\
&\longrightarrow SU(3)_c \times U(1)_{em}.
\end{split}
\label{chichain}
\end{equation}
The first two breakings are achieved by the VEVs of an $SO(10)$ Higgs 45-plet along the $SU(5)$ singlet and $SU(5)$ 24-plet directions respectively. Consider the generation of the magnetic monopole in the first step. To find the structure of this monopole we must first identify the common elements of $SU(5)$ and $U(1)_{\chi}$. Throughout this paper, we normalize the generators of the $U(1)$'s as in Ref.~\cite{Slansky:1981yr}, so that they have the minimal integer charges compatible with a period of $2 \pi$. Under $SU(5)\supset SU(3)_c \times SU(2)_L \times U(1)_Y$, the $\overline{5}$ and $10$ representations of $SU(5)$ decompose as follows:
\begin{equation}
  \overline{5}= (1,2)(-3)+(\overline{3},1)(2),
\label{5barSU5}
\end{equation}
\begin{equation}
  10= (1,1)(6)+(\overline{3},1)(-4)+(3,2)(1).
\label{10SU5}
\end{equation}

\noindent Under the decomposition $SO(10)\supset SU(5) \times U(1)_\chi$, the spinor representation splits as follows:
\begin{equation}
  16= 1(-5)+\overline{5}(3)+10(-1).
\label{16_SO10SU5}
\end{equation}
All group elements that act identically on $16$ act in the same way on all the representations of $SO(10)$ and are thus indistinguishable. 

The $\Z_5$ center of $SU(5)$ lies in $U(1)_Y$ and so its elements will have the form $e^{iY\theta}$, which acts on $\overline{5}$ as indicated below:
\begin{equation}
  \overline{5}= \underbrace{(1,2)(-3)}_{e^{-i3\theta}}+\underbrace{(\overline{3},1)(2)}_{e^{i2\theta}}.
\label{5barSU5underbraceU1Y}
\end{equation}
Since the elements of the center have to be proportional to unity, we must set $3\theta+2\theta=2\pi k$, where k is an integer. The elements of $\Z_5$ are then $e^{i2\pi Y k/5}$ and are generated by $e^{i2\pi Y /5}$. With $SU(5)$ embedded in $SO(10)$, $e^{i2\pi Y /5}$ acts on 16 as:
\begin{equation}
    16= \underbrace{1(-5)}_{1} + \underbrace{\overline{5}(3)}_{e^{i 4\pi/5}} + \underbrace{10(-1)}_{e^{i2\pi/5}}.
    \label{Yact16}
\end{equation}
On the other hand, the $\Z_5$ subgroup of $U(1)_\chi$ is generated by $e^{i2\pi\;\chi/5}$, which acts on the $16$ as follows:
\begin{equation}
     16= \underbrace{1(-5)}_{1} + \underbrace{\overline{5}(3)}_{e^{-i 4\pi/5}} + \underbrace{10(-1)}_{e^{-i2\pi/5}}.
     \label{chiact16}
\end{equation}
Comparing Eqs.~(\ref{Yact16}) and (\ref{chiact16}), we find that:
\begin{equation}
    e^{i\frac{2\pi}{5}\chi}\equiv e^{-i\frac{2\pi}{5}Y}.
    \label{comparYact16chiact16}
\end{equation}

Starting from the unit element, we can then move along $U(1)_\chi$ in the positive direction until the generator of its $\Z_5$ subgroup, which coincides with the inverse generator of the center of $SU(5)$, and return to unity through $SU(5)$. This is a non-contractible loop in the unbroken subgroup $SU(5) \times U(1)_{\chi}$ and yields a monopole that carries both $U(1)_\chi$ and $SU(5)$ magnetic fluxes. Note that in $SU(5)$ we start from the inverse generator of its center but move backwards, and so the flux corresponds to the generator of its center. Returning to Eq.~(\ref{chichain}), the $SU(5)$ breaking in it yields the GUT monopole that carries one unit of Dirac magnetic charge as well as screened color magnetic charge \cite{Daniel:1979yz,Dokos:1979vu}. 

\subsection{Dumbbell configuration}

\begin{figure}[h!]
\centering
      \includegraphics[width=0.5\textwidth,angle=0]{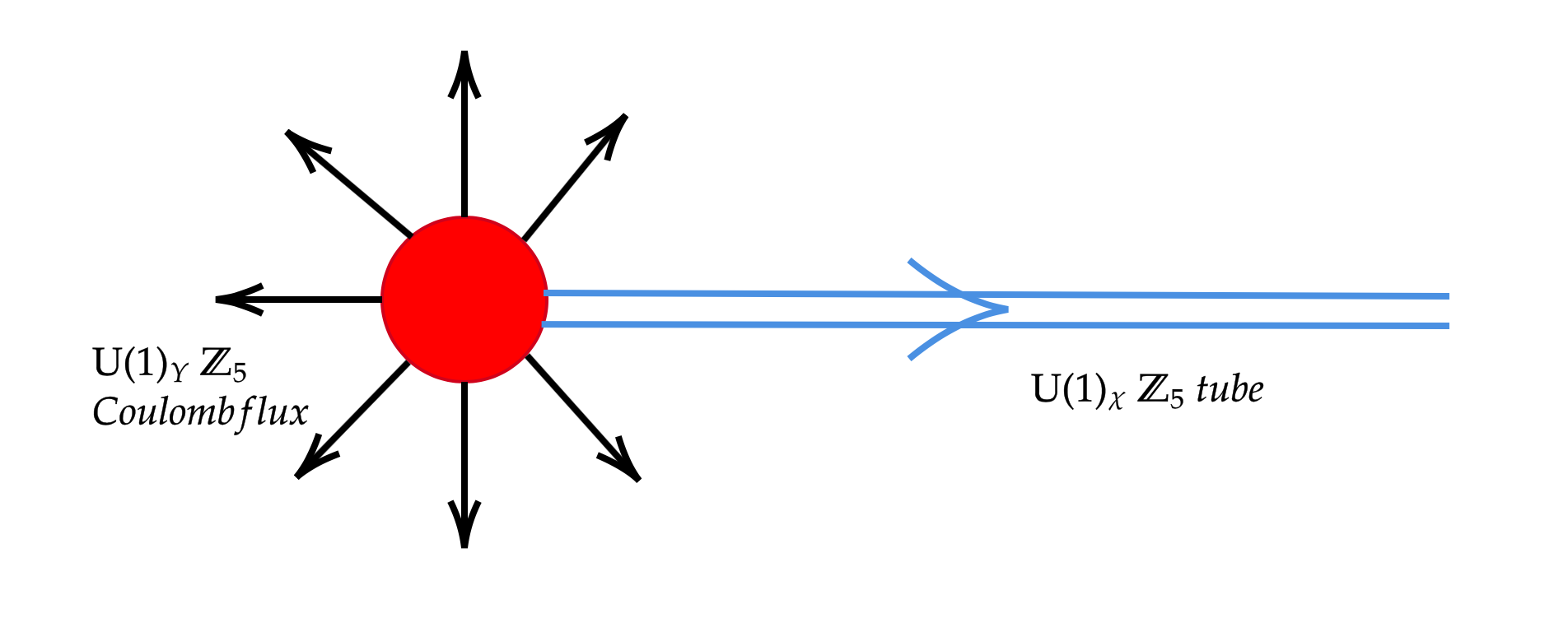}
        \caption {$SO(10)$ monopole carrying $U(1)_Y$ $\Z_5$ Coulomb flux and $U(1)_\chi$ $\Z_5$ magnetic flux tube.}
  \label{coulombtubeSU5}
\end{figure}

As shown above, the first breaking in Eq.~(\ref{chichain}) produces a minimal monopole that carries both $U(1)_\chi$ and $SU(5)$ magnetic fluxes. Indeed, this monopole carries a $U(1)_{\chi}$ magnetic charge associated with a $2\pi/5$ rotation along $U(1)_\chi$ and a magnetic charge associated with a rotation in $SU(5)$ from the unit element to the generator of its center. During the second step of breaking, which is achieved by the VEV of an $SU(5)$ 24-plet Higgs in an $SO(10)$ 45-plet, the $SO(10)$ monopole produced at the first step ends up with a combination of $U(1)_\chi$ and $U(1)_Y$ Coulomb magnetic fields. After the third step of breaking by the VEV of an $SO(10)$ Higgs 16-plet along its $\nu^c$-type component and its conjugate, the $U(1)_\chi$ flux of the $SO(10)$  monopole is squeezed in a tube as shown in \Fig{coulombtubeSU5}. 
We call this tube a $U(1)_\chi$ $\Z_5$ tube to indicate that it corresponds to a $2\pi/5$ rotation along $U(1)_\chi$, but we should keep in mind that as it is generated by the breaking of $U(1)_\chi$ to its $\Z_5$ subgroup, it is not a $\Z_5$ string but rather a $\Z$ string with all its multiples being different. As we go around this tube in the positive direction, the phase of the VEV of the $\nu^c$-type Higgs field changes by $-2\pi$. The $U(1)_Y$ flux of the monopole will be referred to as $U(1)_Y$ $\Z_5$ Coulomb flux to indicate that it corresponds to a $2\pi/5$ rotation along $U(1)_Y$.  

\begin{figure}[h]
\centering
      \includegraphics[width=0.5\textwidth,angle=0]{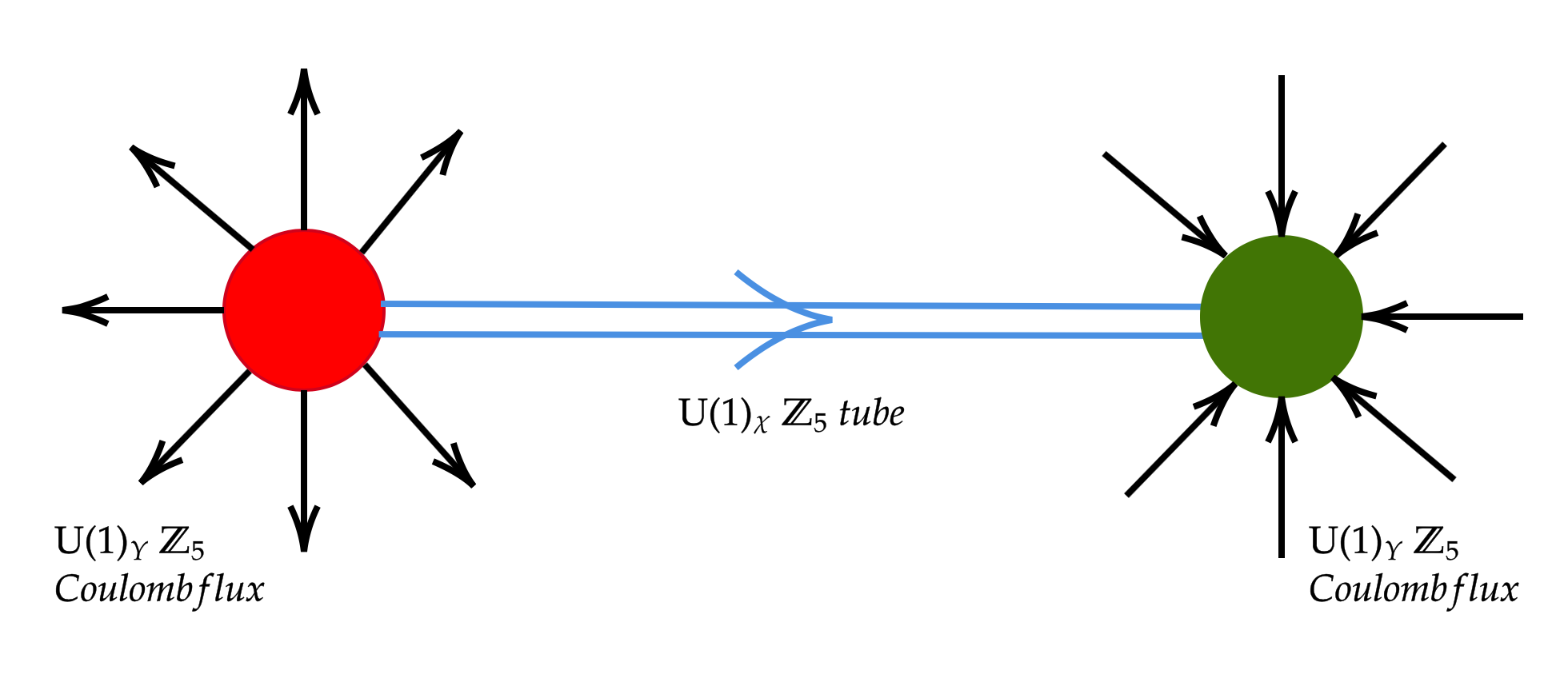}
        \caption{Dumbbell consisting of an $SO(10)$ monopole (red)-antimonopole (green) pair connected by a $U(1)_\chi$ $\Z_5$ flux tube.}
  \label{dumbbellSU5}
\end{figure}

An $SO(10)$ monopole (red) and its antimonopole (green) can be connected by a $U(1)_\chi$ $\Z_5$ tube to form a dumbbell configuration as depicted in \Fig{dumbbellSU5}.   

\noindent The last step in the symmetry breaking in Eq.~(\ref{chichain}) is achieved by the VEVs of the EW Higgs doublets $h_u$, $h_d$ belonging to an $SO(10)$ Higgs $10$-plet and a $\overline{126}$-plet, which under $SU(5)\times U(1)_\chi$ are decomposed as follows:
\begin{equation}
    \begin{split}
     10 &= 5(2)+\overline{5}(-2) \\
    \overline{126} &= 5(2)+\overline{45}(-2)+...
    \end{split}
\label{10and126barSU5chi}
\end{equation}
Under $G_{SM}$, the EW Higgs fields are then identified as:
\begin{equation}
    \begin{split}
    5&=\underbrace{(1,2)(3)}_{h_u}+...,\\
    \overline{5}&=\underbrace{(1,2)(-3)}_{h_d}+...,\\
    \overline{45}&=\underbrace{(1,2)(-3)}_{h_d}+...
    \end{split}
\label{5bar45barGSM}
\end{equation}
\begin{figure}[h]
\centering   

\includegraphics[width=0.4\textwidth,angle=0]{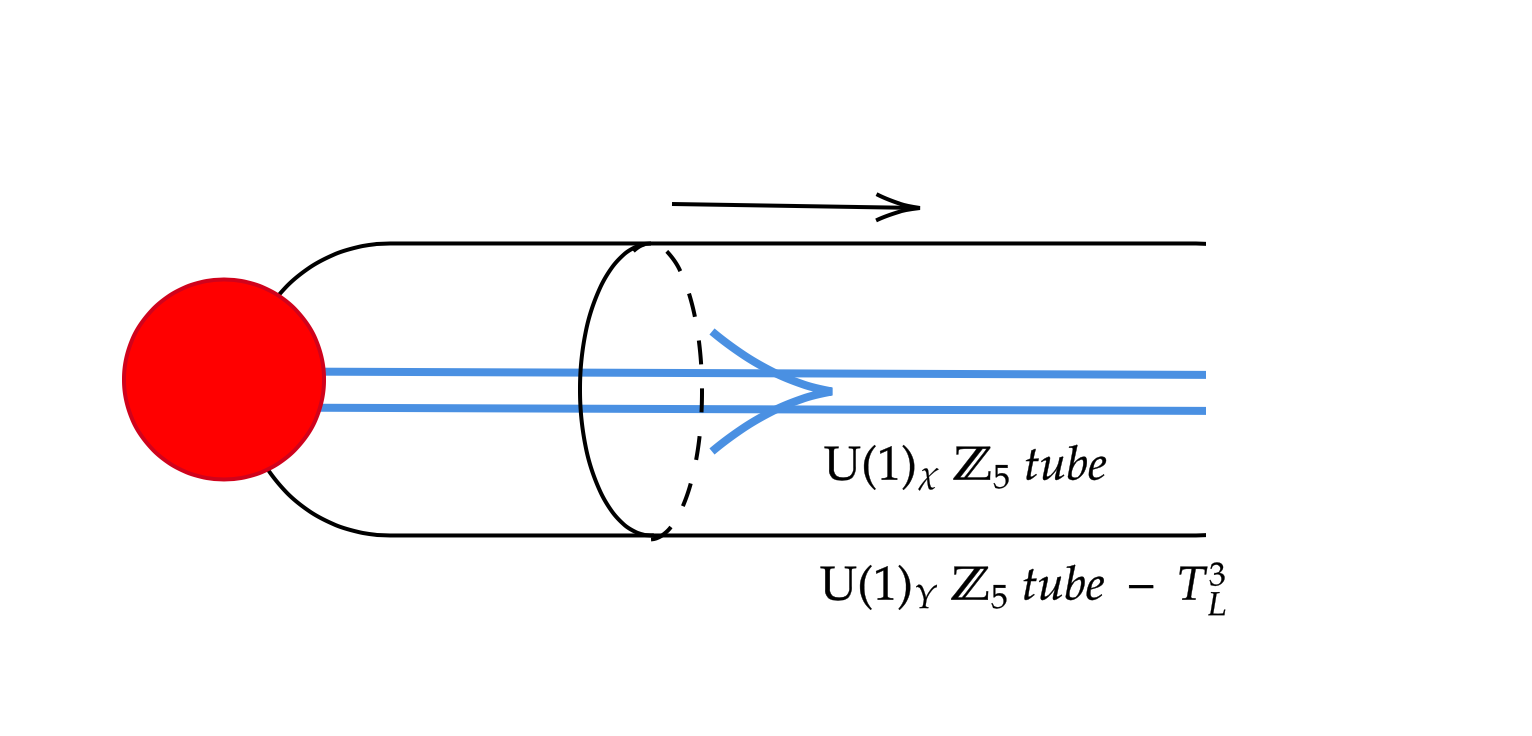}

\includegraphics[width=0.4\textwidth,angle=0]{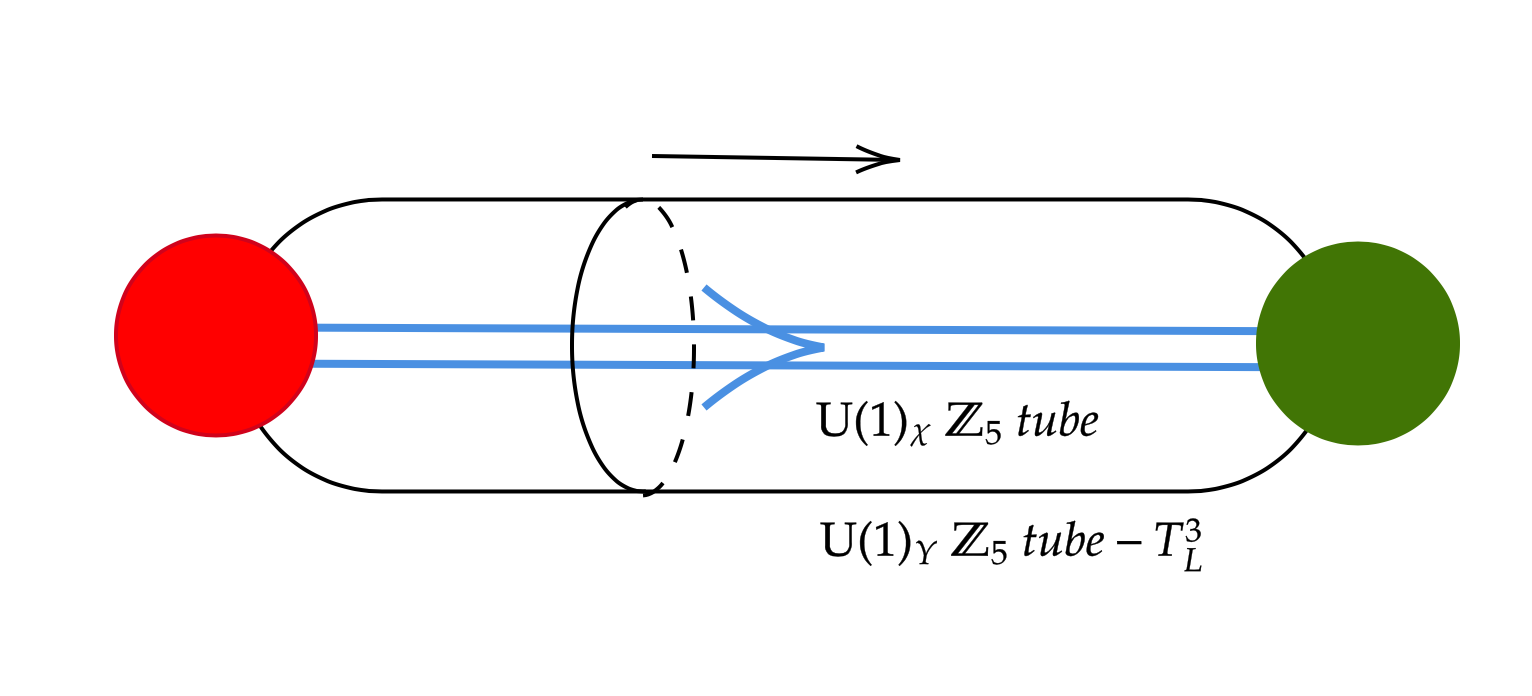}

\caption{$SO(10)$ monopole and dumbbell configurations after the EW breaking.}
\label{monopoledumbbell_afterEW_SU5}
\end{figure}
Before the EW breaking it is legitimate to add a $-T_L^3$ Coulomb flux around the monopole since it corresponds to a contractible loop at this stage (see Ref.~\cite{Lazarides:2021tua}). Note that here we adopt the normalization $T^3_L={\rm diag}(1, -1)$. After the EW breaking, the total Coulomb flux of the monopole is squeezed in an EW tube along the superheavy $U(1)_\chi$ $\Z_5$ flux tube. As one can easily see, the magnetic flux of the EW tube is in the broken direction orthogonal to the electric charge operator $Q$. The change in the phases of the VEVs of $h_u,\;h_d$ as we go around the combined tube are $4\pi,\;-4\pi$ respectively. The phase of the VEV of the $\nu^c$-type Higgs field is obviously not affected by the EW tube and it still changes around the combined tube by $-2\pi$. In \Fig{monopoledumbbell_afterEW_SU5}, we depict the $SO(10)$ monopole and the dumbbell configuration with the combined tube after the EW breaking.

\subsection{Starfish configuration}
If we break $SU(5)\longrightarrow G_{SM}$ at the first step simultaneously with the breaking of $SO(10)\longrightarrow SU(5)\times U(1)_\chi$, we obtain, of course, the standard $SU(5)$ monopole and also the $SO(10)$ monopole. 
\begin{figure}[h]
\centering
      \includegraphics[width=0.55\textwidth,angle=0]{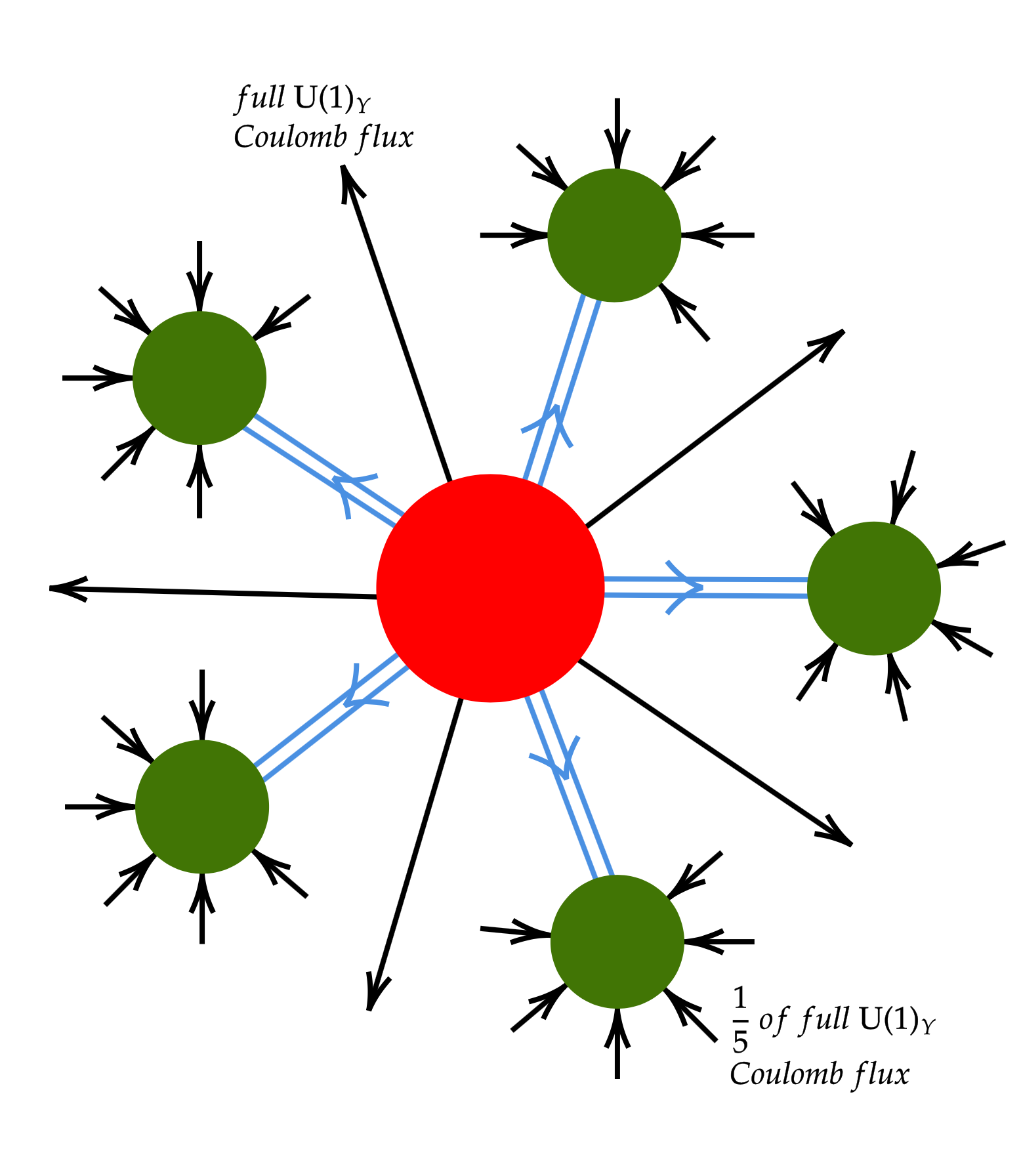}
        \caption{Starfish-like configuration with a central multimonopole connected to five $SO(10)$ antimonopoles by $U(1)\chi$ $\Z_5$ tubes. For simplicity, here and in all the subsequent figures we display the Coulomb flux of (anti)monopoles only once.}
  \label{starfishSU5}
\end{figure}
Taking five $SO(10)$ monopoles together we get a multimonopole configuration with full $U(1)_\chi$ and $U(1)_Y$ Coulomb fluxes corresponding to $2\pi$ rotations about these two abelian groups. 
\begin{figure}[h]
\centering
      \includegraphics[width=0.55\textwidth,angle=0]{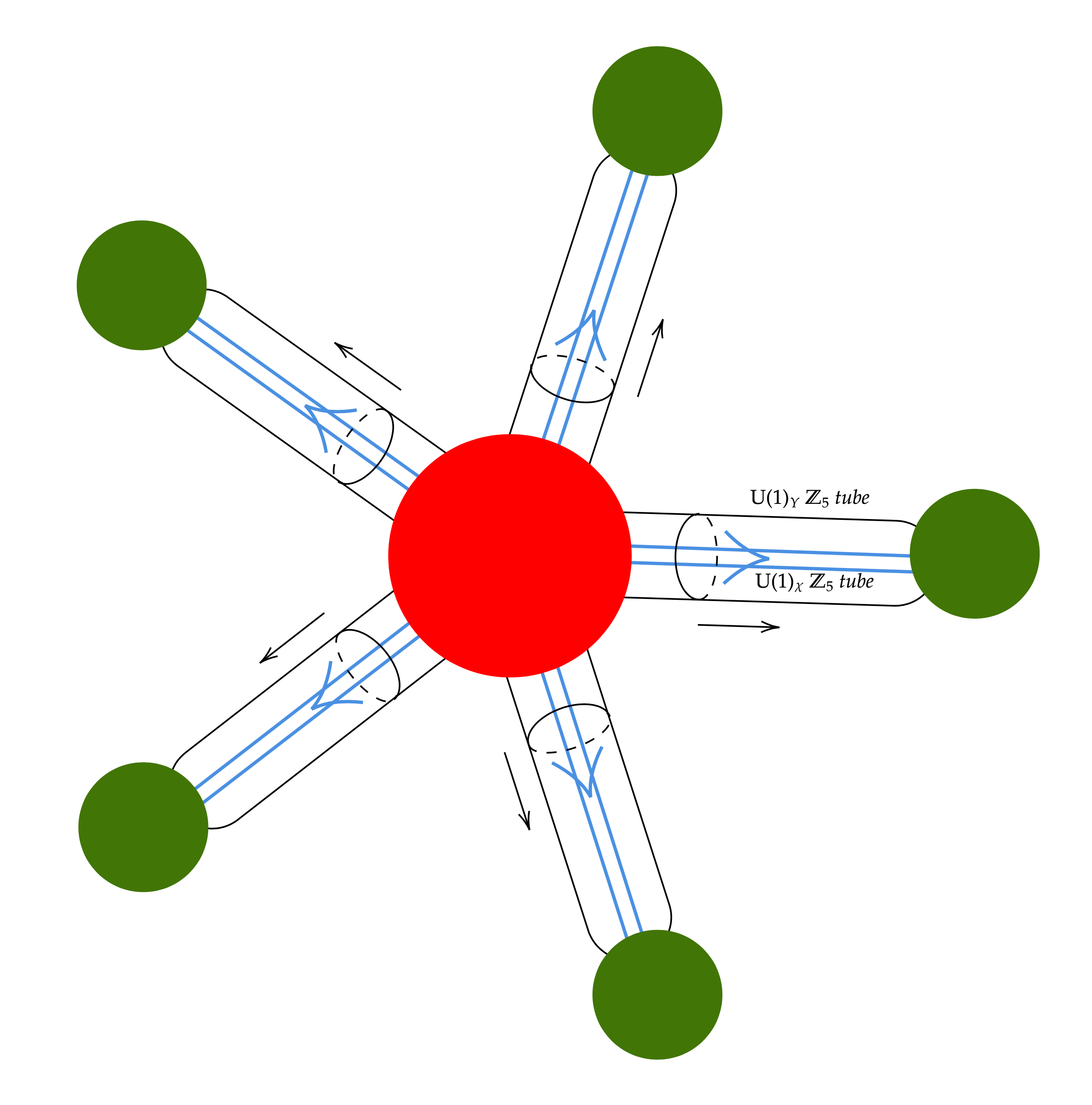}
        \caption{Squeezing of the $U(1)_Y$ flux of the central multimonopole of the starfish configuration into five $U(1)_Y\; \Z_5$ tubes after the EW. These tubes meet the corresponding tubes from the antimonopoles. Here and in all figures, the tube fluxes are displayed only once.}
  \label{starfish_afterEW_SU5}
\end{figure}
This multimonopole configuration may be stable for a suitable choice of parameters (compare with Refs.~\cite{Gardner:1983uu,Vachaspati:1995yp}, where the stability of the $SU(5)$ monopole with magnetic charge five times the Dirac charge is discussed). After the breaking of $U(1)_\chi$, its flux is squeezed in five $U(1)_\chi$ $\Z_5$ tubes which can terminate on five $SO(10)$ antimonopoles, thereby forming a starfish-like configuration shown in \Fig{starfishSU5}.
After the EW breaking, the $U(1)_Y$ flux of the central multimonopole is also squeezed in five $U(1)_Y \Z_5$ tubes which meet the analogous tubes from the antimonopoles (see \Fig{starfish_afterEW_SU5}).

Next let us consider the case where $SO(10)$ breaks via $SU(5)\times U(1)_\chi$ as in Eq.~(\ref{chichain}).
For $SU(5)$ unbroken the full $U(1)_Y$ Coulomb flux can be removed from the multimonopole in 
\Fig{starfishSU5} since $U(1)_Y$ is contractible in $SU(5)$. This makes the stability of the central
multimonopole much more secure in a range of parameters since it reduces its energy below the energy of five single monopoles. However, after the EW breaking the $U(1)_Y$ $\Z_5$ tubes from the peripheral antimonopoles will be created and terminate on the central multimonopole. This requires that the $U(1)_Y$ flux of this monopole is excited again and this may destabilize it leading to its breaking into five monopoles. The whole starfish-like configuration may then decay, after the EW breaking, into five dumbbells which finally contract.

\subsection{Necklace configuration}

\begin{figure}[h!]
\centering
      \includegraphics[width=0.6\textwidth,angle=0]{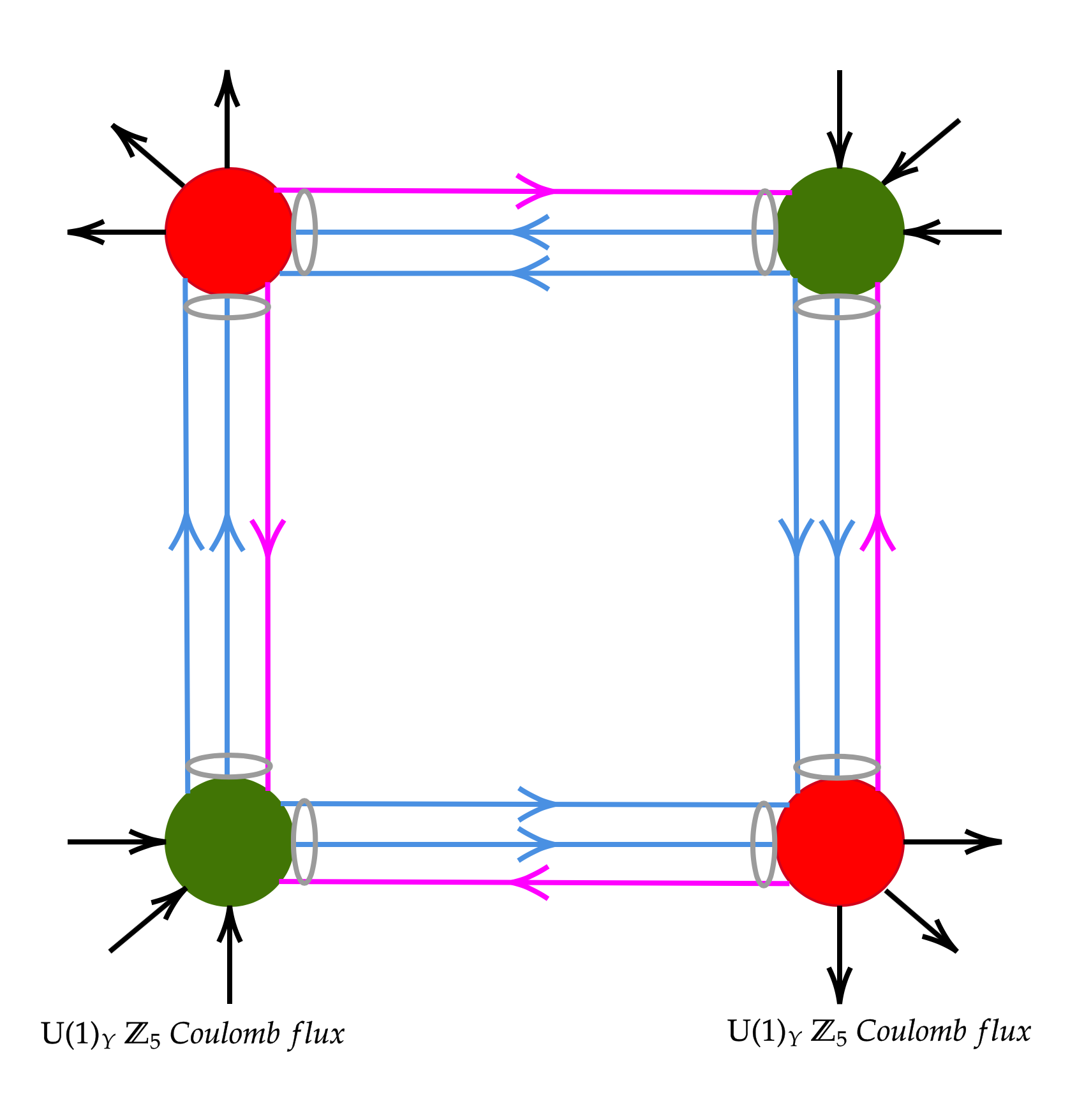}
        \caption{Necklace of monopoles (red) and antimonopoles (green) connected by $U(1)_\chi$ $\Z_{10}$ tubes, which carry half the flux of the $\Z_5$ tubes and correspond to a rotation by $2\pi/10$ along $U(1)_\chi$. These tubes can be thought of as hybrid structures consisting of a $\Z_2$ tube (magenta) and two $\Z_5$ anti-tubes (blue). The tubes are all $\Z$ strings and their accompanying index refers only to the flux that they carry. The real $\Z_2$ string is the necklace itself in the sense that if one inverts all its arrows it coincides with itself.}
  \label{necklaceSU5}
\end{figure}

We now consider the case where the third breaking in Eq.~(\ref{chichain}) is performed by the VEV of the $\nu^c\nu^c$-type component in a Higgs 126-plet of $SO(10)$ and its conjugate. In this case, the $\Z_2$ subgroup of the center of $SO(10)$ remains unbroken, which leads to the generation of $\Z_2$ strings as shown in Ref.~\cite{Kibble:1982ae}. The group $U(1)_\chi$ is broken to its $\Z_{10}=\Z_5\times \Z_2$ subgroup ($\Z_5$ is the center of $SU(5)$), and thus the flux tube in \Fig{coulombtubeSU5} splits into two equivalent flux tubes emerging from the monopole, as shown in \Fig{necklaceSU5}. We can think of this $U(1)_\chi$ $\Z_{10}$ flux tube as consisting of one $\Z_2$ tube and two $\Z_5$ anti-tubes since $1/10=1/2-2/5$. However, we note that all these tubes are really $\Z$ strings as they are generated by the breaking of $U(1)_\chi$ to discrete subgroups. We call them $U(1)_\chi$ $\Z_n$ tubes just to indicate that they correspond to a rotation by $2\pi/n$ along $U(1)_\chi$ in the positive direction. In particular, we orient even the $\Z_2$ tubes to indicate that the rotation is in the positive direction, in contrast to the $\Z_2$ strings which are not oriented.  
Connecting monopoles and antimonopoles by $U(1)_\chi$ $\Z_{10}$ tubes, we can form necklaces (see \Fig{necklaceSU5}), which are the real $\Z_2$ strings. Indeed, inverting all the arrows in a necklace we obtain an equivalent necklace. This is clear since topologically stable $\Z_2$ strings are generated if we break $SO(10)$ to $G_{SM}\times \Z_2$ and not just $U(1)_\chi$ to $\Z_2$. Consequently, the scale of the $SO(10)$ breaking must appear in the strings and it does so via the $SO(10)$ monopoles and antimonopoles which participate in the necklace as beads. Going around one of the tube segments connecting monopoles and antimonopoles in \Fig{necklaceSU5} in the positive direction with respect to the orientation of its $\Z_2$ tube, the phase of the VEV of the $\nu^c\nu^c$-type Higgs field changes by $-2\pi$.

\begin{figure}[h]
\centering
      \includegraphics[width=0.65\textwidth,angle=0]{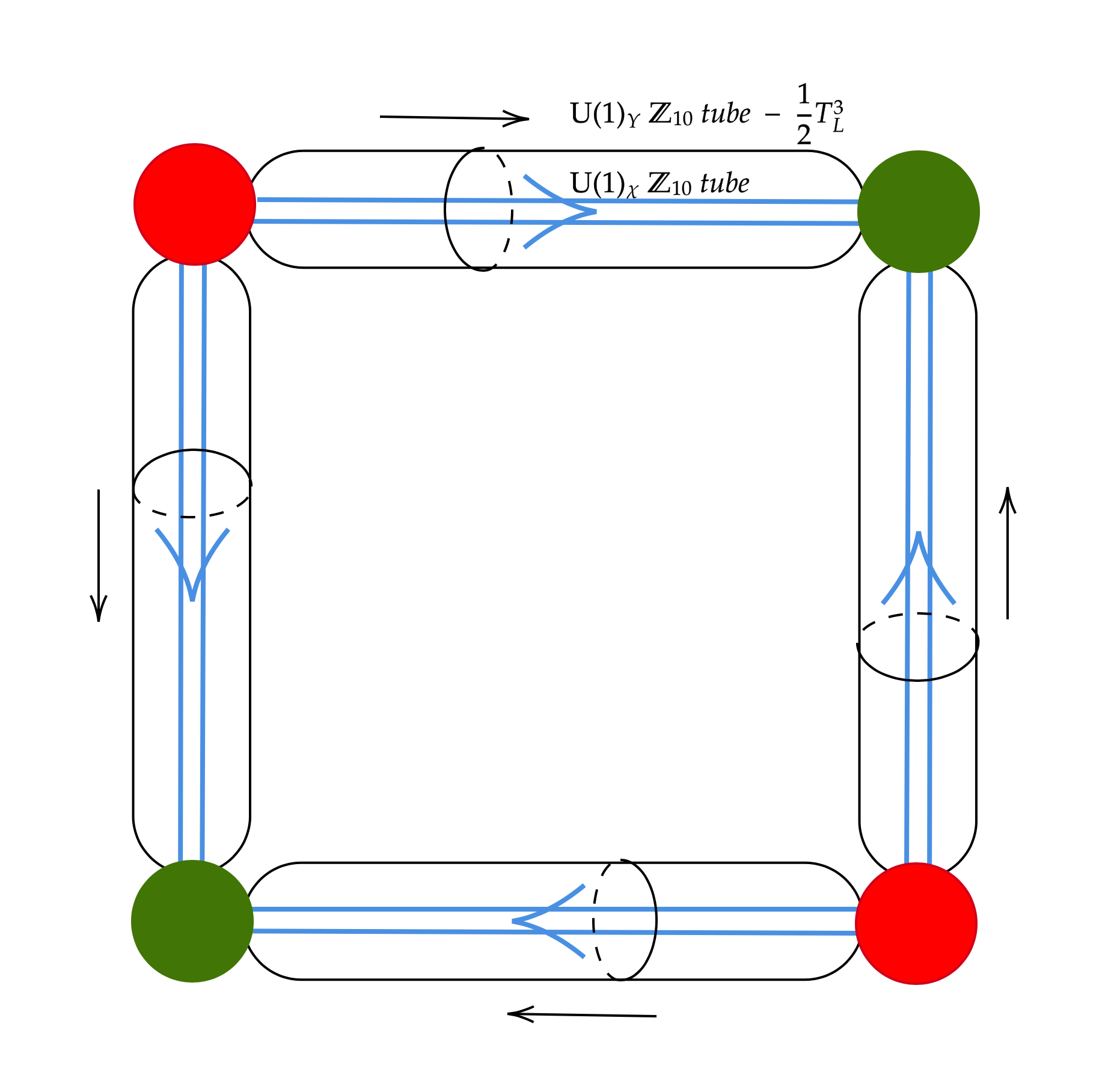}
        \caption{Necklace configuration after EW breaking. For simplicity we do not show here the structure of $U(1)_\chi \; \Z_{10}$ tubes.}
  \label{necklace_afterEW_SU5}
\end{figure}

Following the EW breaking, the EW flux tubes are formed around the $U(1)_\chi$ $\Z_{10}$ tubes of the necklace, which takes the form shown in \Fig{necklace_afterEW_SU5}. These tubes carry half the flux of the EW tube in \Fig{monopoledumbbell_afterEW_SU5}. As we go around the combined tubes, the phases of the VEVs of $h_u$, $h_d$ change by $2\pi$, $-2\pi$ respectively, and the phase of the VEV of the $\nu^c\nu^c$-type Higgs field changes by $-2\pi$. Inflating away the monopoles, one can generate a network of $\Z$ strings, which though have the tendency to create monopole-antimonopole pairs on them and become $\Z_2$ strings. Depending on the breaking scales associated with the monopoles and the strings, the $\Z$ strings can be metastable or even quasistable against monopole-antimonopole pair creation.                 

\subsection{Walls bounded by Necklaces}
\label{subsec:Wall_bounded_Necklaces}
Before discussing the formation of walls bounded by necklaces in the $SO(10)$ model, it is instructive to consider a simple example provided by the symmetry breaking $SU(2) \xrightarrow{} U(1)\xrightarrow{} \Z_2\xrightarrow{} \{e\}$, where $e$ denotes the identity element of $SU(2)$. The first step in the breaking produces a monopole with flux corresponding to a $2 \pi$ rotation about $T^3={\rm diag}(1, -1)$. Put differently, the monopole carries two units of Dirac magnetic charge and is the well known 't Hooft-Polyakov monopole \cite{tHooft:1974kcl,Polyakov:1974ek}. The breaking of $U(1)$ to $\Z_2$ produces $\Z$ strings corresponding to a half rotation about $T^3$, i.e. from zero to $\pi$. The flux of the monopole, since it corresponds to a full rotation about $T^3$, forms two tubes and thus necklaces can appear, which are $\Z_2$ strings. The final spontaneous breaking of $\Z_2$ means that these strings have to disappear, which they do by forming boundaries of $\Z_2$ walls. 

\begin{figure}[h]
\centering
      \includegraphics[width=0.65\textwidth,angle=0]{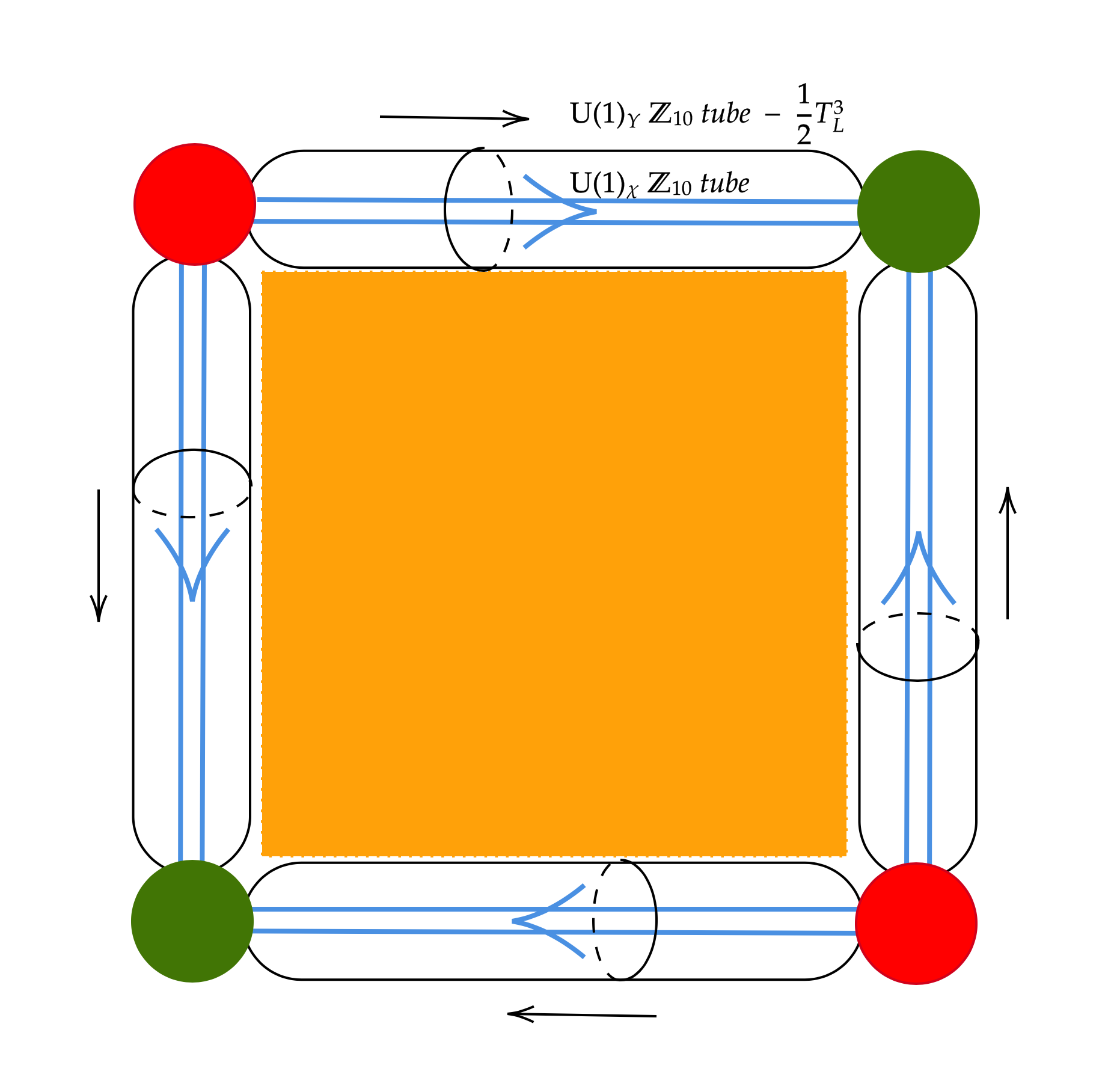}
        \caption{Emergence of a $\Z_2$ domain wall (orange) from each segment of the necklace due to the breaking of the $\Z_2$ subgroup of $U(1)_{\chi}$ by the VEV of a $\nu^c$-type Higgs field and its conjugate, with the necklace ultimately becoming the boundary of the $\Z_2$ wall.}
  \label{domainwall_necklace_SU5}
\end{figure}

In the $SO(10)$ model under discussion, similar structures would arise in a number of ways. Recall that we have identified $\Z_2$ strings which are necklaces of monopoles and antimonopoles. If this $\Z_2$ is subsequently broken, say by the VEV of a Higgs 16-plet and its conjugate, we obtain structures that we call “walls bounded by necklaces”. Let us consider the breaking of the $\Z_2$ subgroup of $U(1)_{\chi}$  by the VEV of a $\nu^c$-type Higgs field and its conjugate. Going around any tube segment in the necklace, the 16-plet VEV and its conjugate change sign. Consequently, a $\Z_2$ domain wall emerges from each segment through which this VEV changes sign. Therefore, the necklace becomes the boundary of a $\Z_2$ wall (shaded square) as shown in \Fig{domainwall_necklace_SU5}.

\section{$\mathbf{SO(10)}$ breaking via $\mathbf{SU(4)_c \times SU(2)_L \times SU(2)_R}$}
\label{sec:422}
\subsection{Dumbbell configuration}
\begin{figure}[h]
\centering

      \includegraphics[width=0.6\textwidth,angle=0]{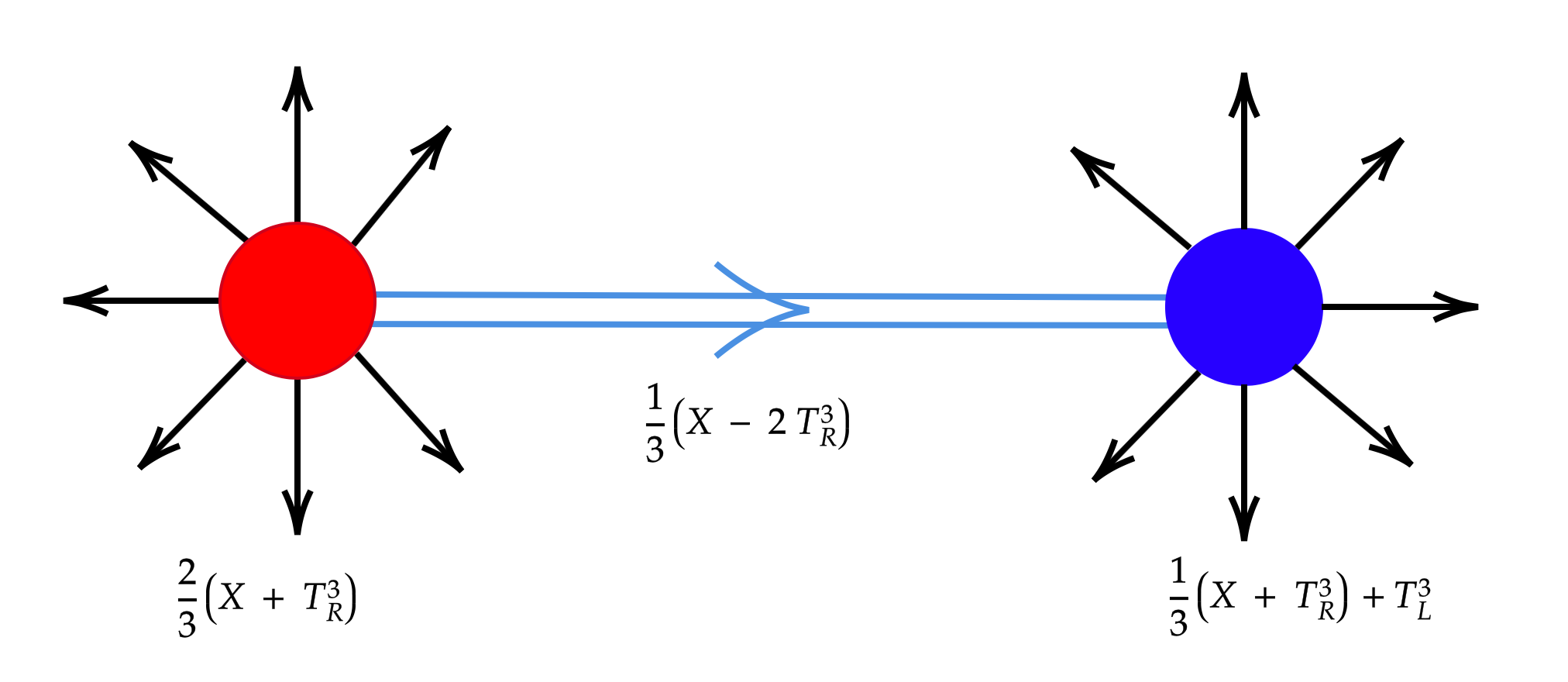}
        \caption{Dumbbell consisting of a red and a blue magnetic monopole from the breakings of $SU(4)_c$ to $SU(3)_c\times U(1)_{B-L}$ and $SU(2)_R$ to $U(1)_R$ respectively connected by a magnetic flux tube generated by the breaking $U(1)_{B-L} \times U(1)_R$. A full $T^3_L$ Coulomb flux is added in the blue monopole.}
  \label{dumbbell422}
\end{figure}

We consider the following SO(10) breaking chain:
\begin{equation}
\begin{split}
    SO(10) &\longrightarrow SU(4)_c \times SU(2)_L \times SU(2)_R\\
    &\longrightarrow
SU(3)_c \times U(1)_{B-L} \times SU(2)_L \times U(1)_R\\
&\longrightarrow SU(3)_c \times SU(2)_L \times U(1)_Y\\
&\longrightarrow SU(3)_c \times U(1)_{em}
\end{split}
\label{422chain}
\end{equation}

\noindent following the discussion of Ref. \cite{Lazarides:2019xai}. The first breaking yields $\Z_2$ monopoles \cite{Lazarides:1980cc,Lazarides:2019xai}, which eventually turn into the superheavy GUT monopole. The second breaking yields two kinds of magnetic monopoles which correspond to the breaking of $SU(4)_c$ to $SU(3)_c \times U(1)_{B-L}$, and $SU(2)_R$ to $U(1)_R$. They were coined red and blue monopoles respectively in Ref.~\cite{Lazarides:2019xai} and correspond to a full rotation about $X=B-L + 2T^8_c/3$ and $T^3_R$, where $B$ and $L$ are the baryon and lepton numbers and $T^8_c={\rm diag}(1, 1, -2)$, $T^3_R={\rm diag}(1, -1)$. The breaking of $U(1)_{B-L} \times U(1)_R$ to $U(1)_Y$ is achieved by the VEVs of a $\nu^c$-type Higgs field from a 16-plet and its conjugate, and this generates a tube that connects a red and a blue monopole leading to the dumbbell configuration in Fig. 1 of Ref.~\cite{Lazarides:2019xai}. In order to find the correct transformation of this dumbbell after the EW breaking, we employ a trick which was introduced in Ref.~\cite{Lazarides:2021tua}. At the level of unbroken $U(1)_{B−L} \times U(1)_R$, the $SU(2)_L$ gauge group is also unbroken. In this case, a full rotation about $T^3_L$ is contractible, and it is legitimate to add a full $T^3_L$ Coulomb flux emerging from the blue monopole. The dumbbell then takes the form shown in \Fig{dumbbell422}.

\begin{figure}[h]
\centering

      \includegraphics[width=0.6\textwidth,angle=0]{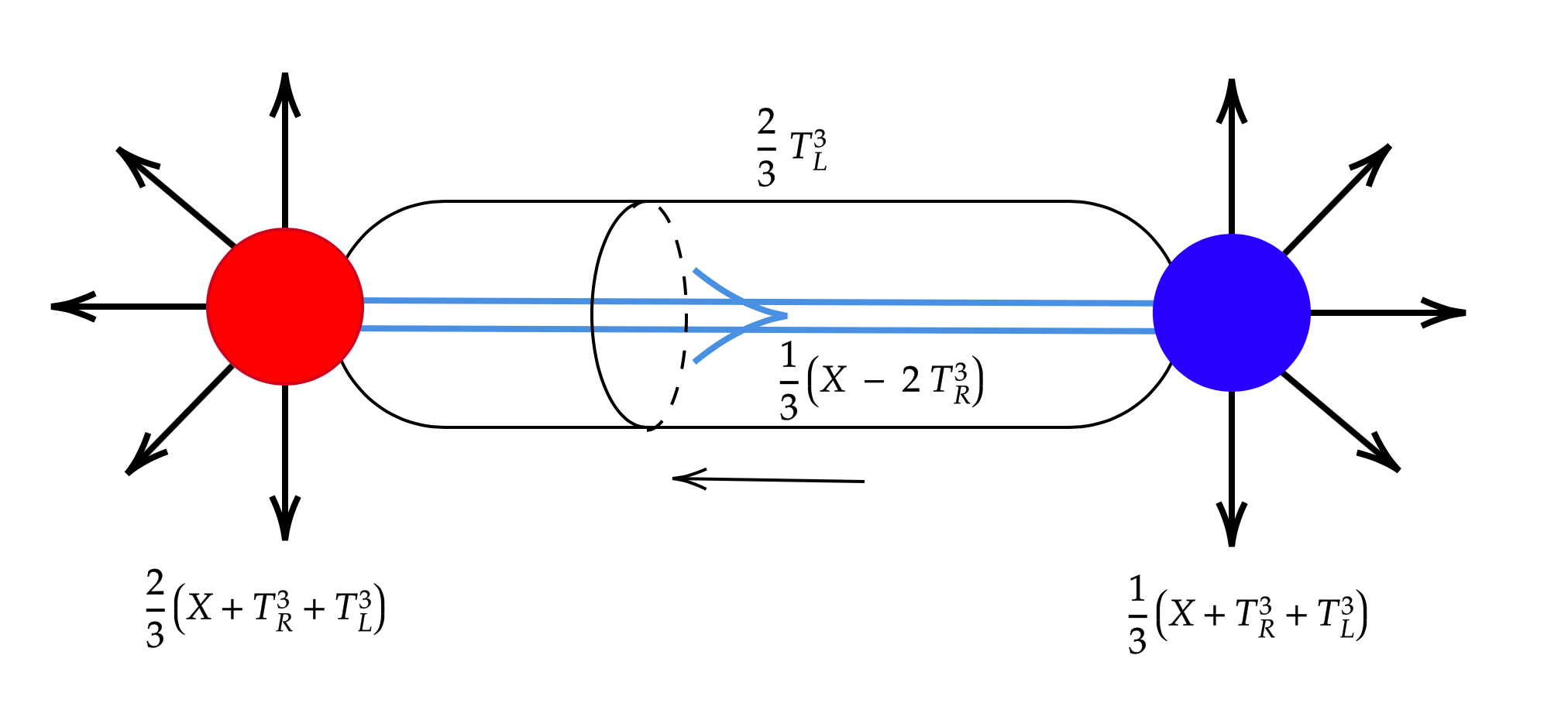}
        \caption{Dumbbell configuration after the EW breaking.}
  \label{dumbbell_afterEW_422}
\end{figure}
Note that the $\nu^c$-type VEV has $X = 1$, $T^3_R= −1$, $T^3_L=0$. Therefore the two Coulomb fluxes in \Fig{dumbbell422} correspond to unbroken generators at this stage. At the EW breaking a tube with flux $-2T^3_L/3$ has to be sent along the superheavy tube from the red to the blue monopole. The Coulomb flux is then $2(X+T^3_R+T^3_L)/3$ for the red monopole, and $(X+T^3_R+T^3_L)/3$ for the blue one. Recall that $X + T^3_R+T^3_L=2Q + 2T^8_c/3$, the Schwinger monopole flux plus some color magnetic flux, as required by the Dirac quantization condition. The transformation of the configuration in \Fig{dumbbell422} after the EW breaking is shown in \Fig{dumbbell_afterEW_422}. Recall that the VEVs of the EW Higgs doublets $h_u$, $h_d$ have $X = 0$, $T^3_R = 1$, $−1$, $T^3_L= −1$, $1$ respectively, and consequently their phases remain constant around the tube. The phase of the $\nu^c$-type Higgs field changes by $2\pi$. This configuration can contract to a Schwinger monopole with Coulomb flux $2Q + 2T^8_c/3$.

\subsection{Polypole configurations}
\begin{figure}[h]
\centering

      \includegraphics[width=0.45\textwidth,angle=0]{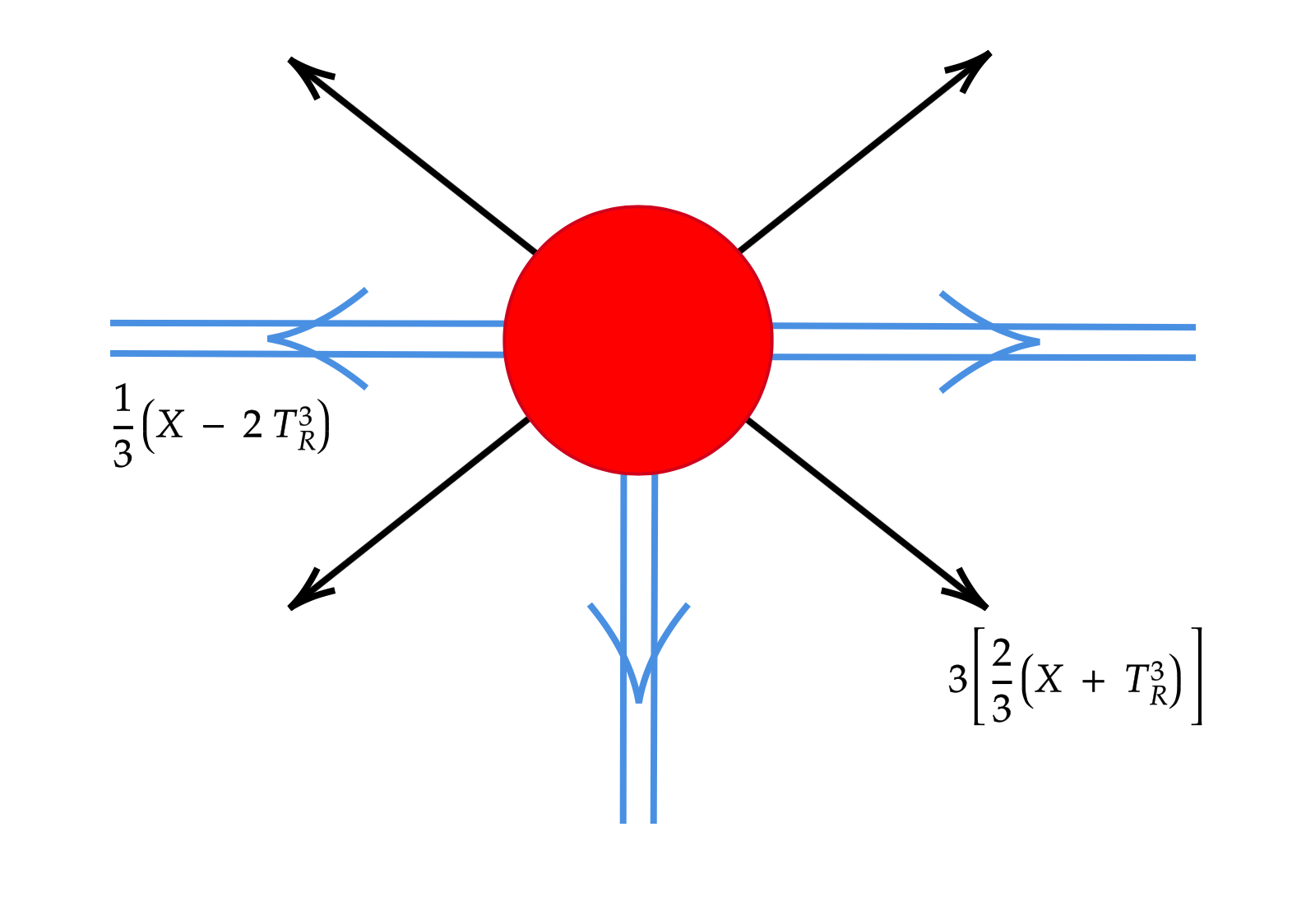}
        \caption{Emergence of three flux tubes after the breaking of $U(1)_{B−L} \times U(1)_R$ from a configuration consisting of three red monopoles.}
  \label{monopole_threetubes_422}
\end{figure}

We now discuss the polypole configurations ensuing from the breaking of $SO(10)$ through $SU(4)_c \times SU(2)_L \times SU(2)_R$. At the $SU(3)_c \times SU(2)_L \times U(1)_{B−L} \times U(1)_R$ level, we take three red monopoles together which presumably can give a stable red trimonopole in a suitable range of the parameters. The total flux of this Coulombic configuration is 3X. After the breaking of $U(1)_{B−L} \times U(1)_R$ to $U(1)_Y$, three tubes emerge from this configuration as shown in \Fig{monopole_threetubes_422} and the remaining Coulomb flux is $3[2(X + T^3_R)/3]$. 
\begin{figure}[h]
\centering

      \includegraphics[width=0.67\textwidth,angle=0]{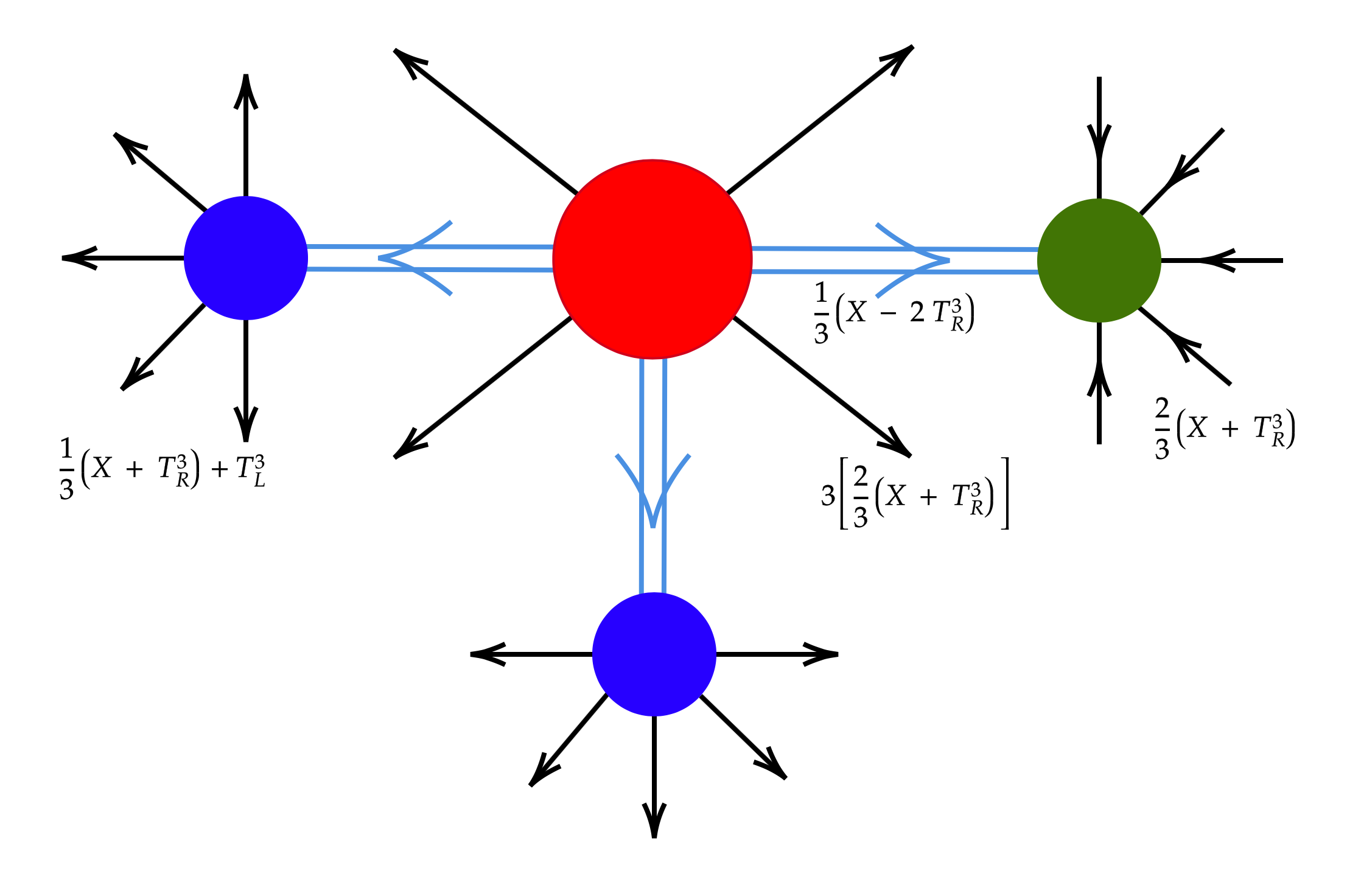}
        \caption{A polypole configuration with the three tubes from the red trimonopole after the breaking of $U(1)_{B−L} \times U(1)_R$ terminating on one green (anti-red) and two blue monopoles.}
  \label{polypole422}
\end{figure}

The tubes can terminate on green (anti-red) monopoles or blue monopoles. An example with one anti-red and two blue monopoles is depicted in \Fig{polypole422}. Note that following Ref.~\cite{Lazarides:2021tua}, an extra $T^3_L$ flux was added in each blue monopole.

\begin{figure}[h]
\centering

      \includegraphics[width=0.8\textwidth,angle=0]{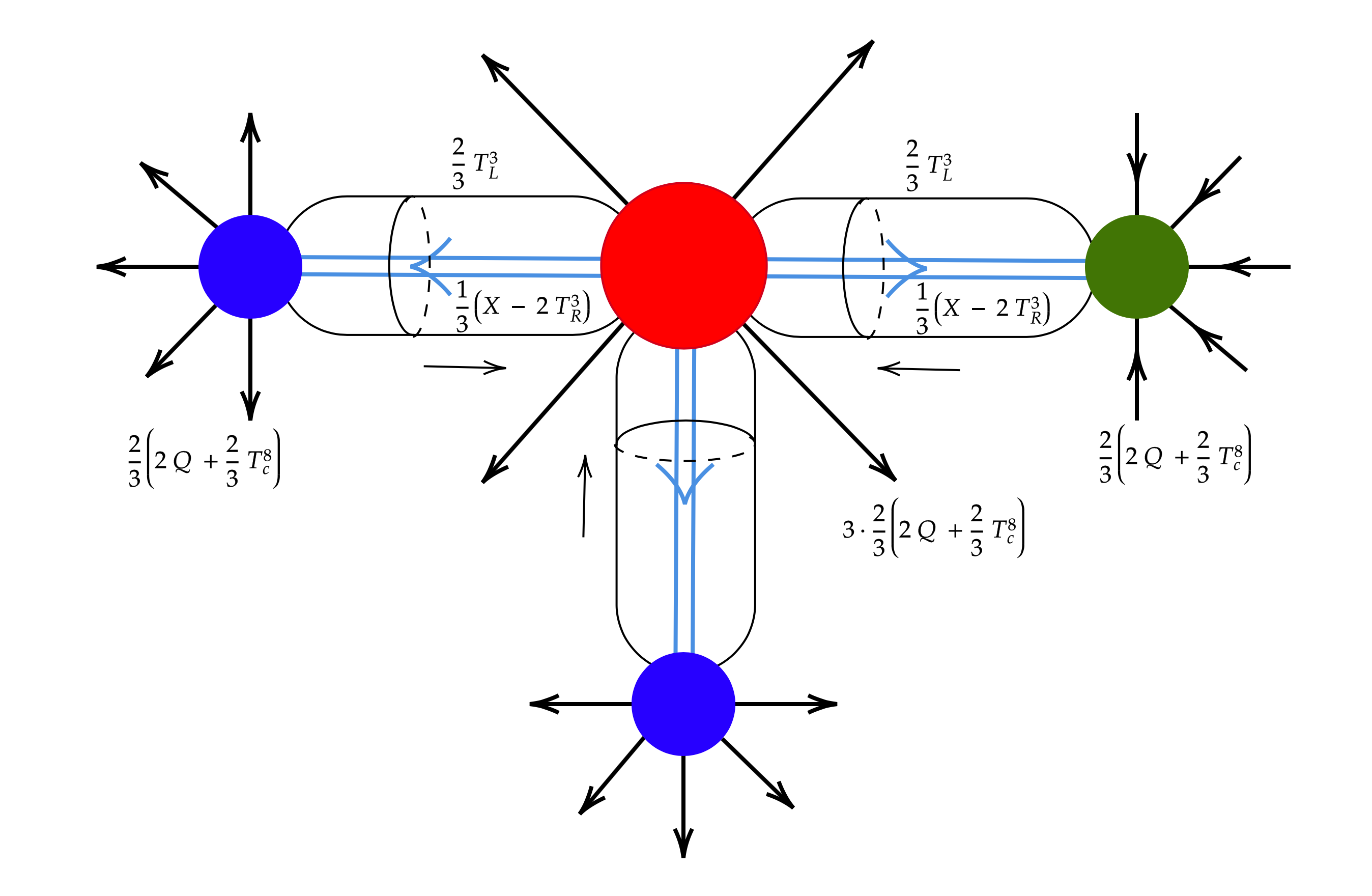}
        \caption{Polypole configuration in \Fig{polypole422} after the EW symmetry breaking.}
  \label{polypole_afterEW_422}
\end{figure}

The overall Coulomb flux of the configuration in \Fig{polypole422} is twice the Schwinger flux, but we can have such configurations with 0, 1, 2, or 3 Schwinger fluxes depending on the number of green (anti-red) and blue monopoles involved. After the EW breaking the configuration in \Fig{polypole422} transforms into the one shown in \Fig{polypole_afterEW_422}.
\begin{figure}[h]
\centering

      \includegraphics[width=0.8\textwidth,angle=0]{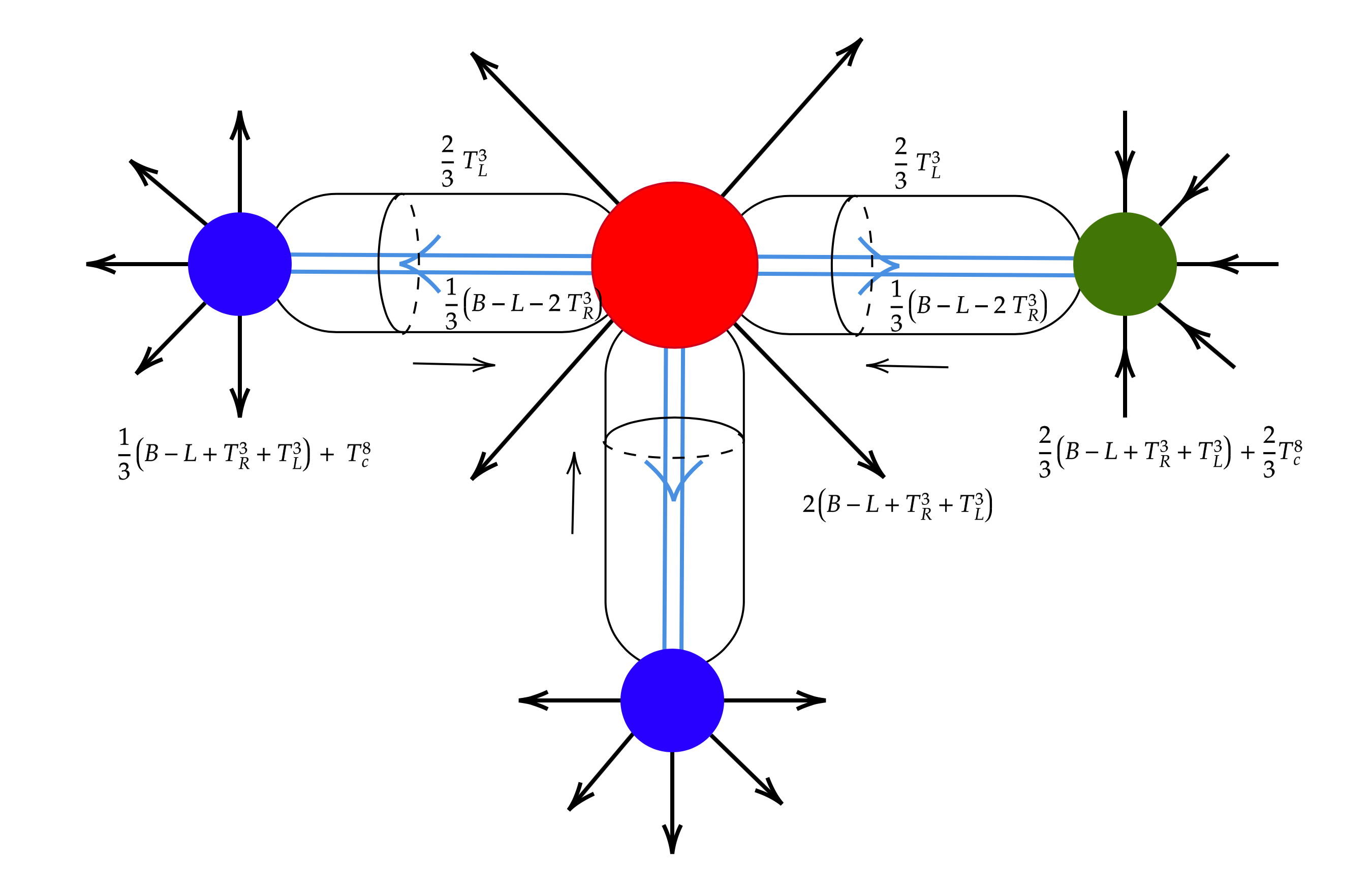}
        \caption{The polypole configuration after removing the contractable loop in $SU(3)_c$ from the central trimonopole to ensure its stability.}
  \label{stablepolypole_afterEW_422}
\end{figure}

To secure the stability of the central trimonopole we can remove the contractable loop in $SU(3)_c$ so that the initial Coulomb flux of the trimonopole is 3(B-L) instead of 3X. This reduces the energy of the trimonopole below the energy of three single red monopoles. The configuration in  \Fig{polypole_afterEW_422} then transforms into the one shown in \Fig{stablepolypole_afterEW_422}. Here we took an extra $T^8_c$ flux together with the extra $T^3_L$ flux with each blue monopole. The Coulomb fluxes correspond to unbroken generators and the behavior of the VEVs of the $\nu^c$-type, $h_u$, $h_d$ Higgs fields around the tube is the same as before. The total Coulomb flux is again twice the Schwinger flux.

\subsection{Necklace configurations}
\begin{figure}[h]
\centering

\includegraphics[width=0.45\textwidth,angle=0]{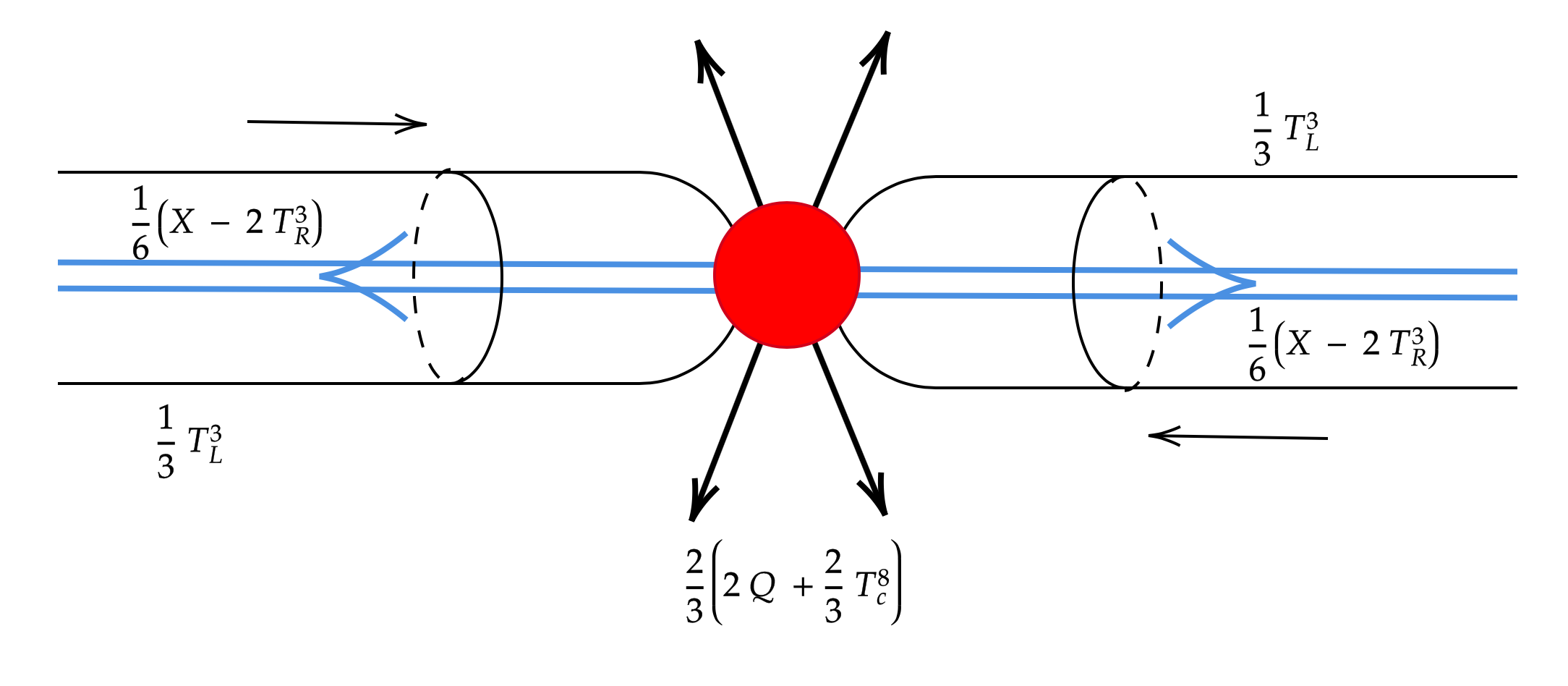}
\includegraphics[width=0.45\textwidth,angle=0]{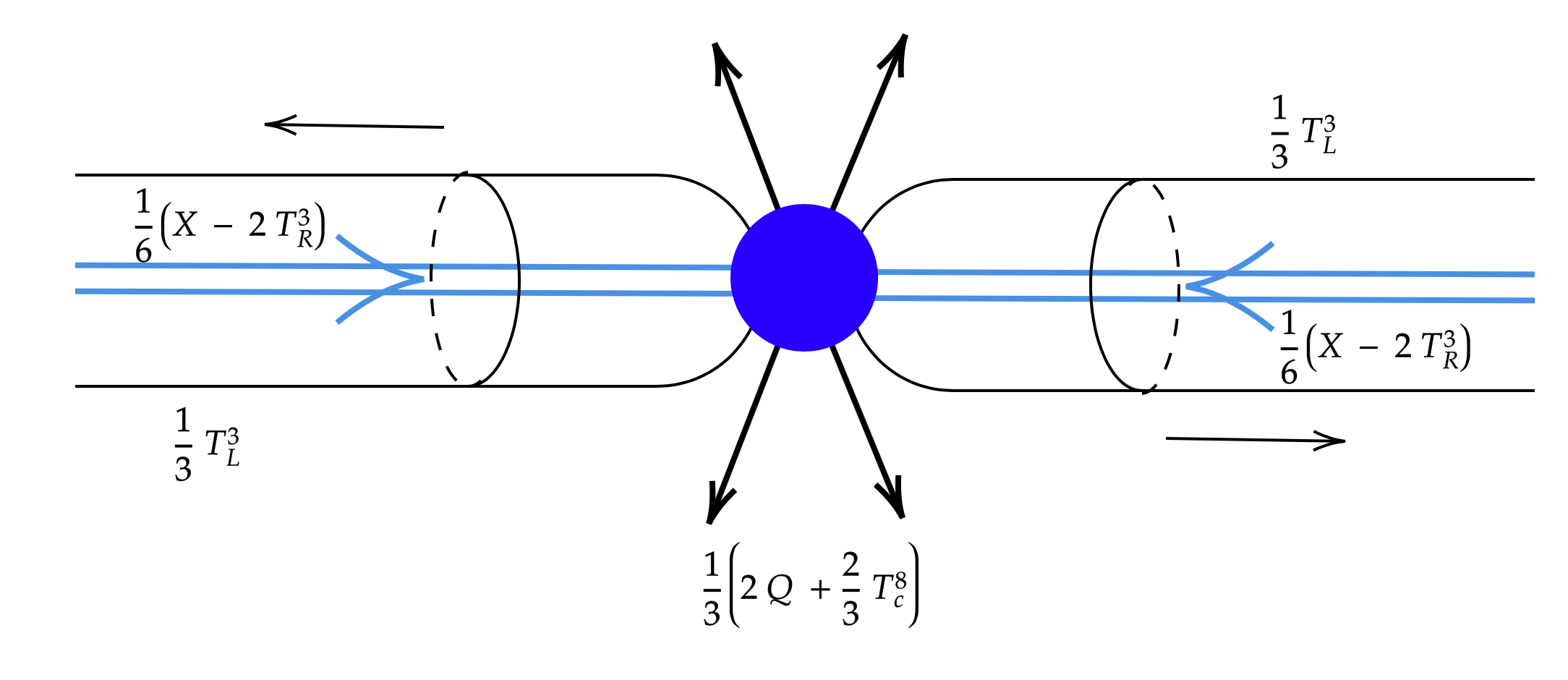}
\caption{The splitting of the combined tube in \Fig{dumbbell_afterEW_422} into two equivalent tubes in the case with topologically stable $\Z_2$ strings.}
  \label{splittube422}
\end{figure}

\begin{figure}[h]
\centering

      \includegraphics[width=0.8\textwidth,angle=0]{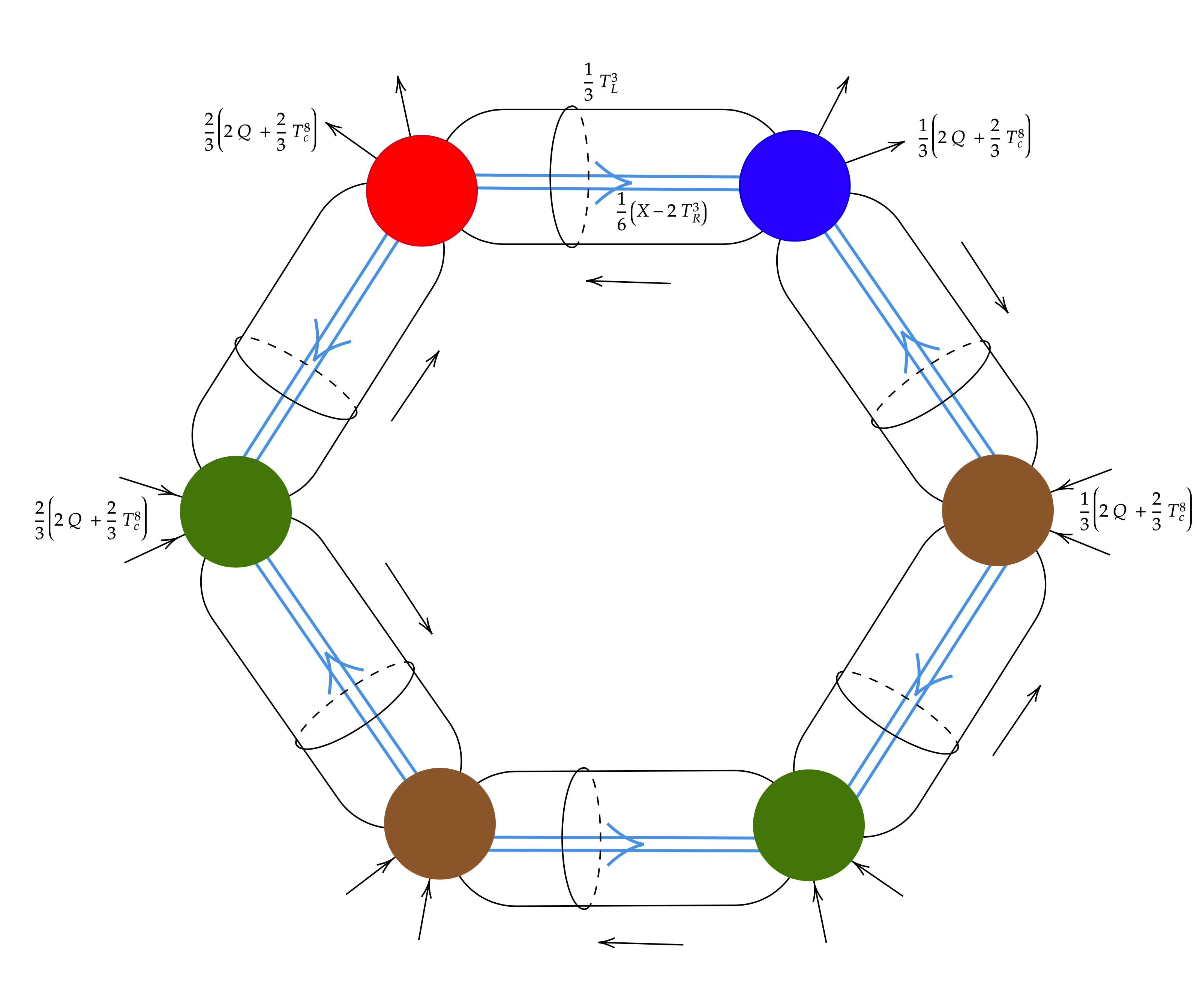}
        \caption{An example of necklace constructed by connecting the loose ends of the flux tubes in \Fig{splittube422}. The antimonopole of the blue monopole is the brown monopole, and the antimonopole of the red monopole is the green monopole. This necklace is a realization of the $\Z_2$ string.}
  \label{necklace422}
\end{figure}

In the case with topologically stable $\Z_2$ strings where the VEVs of a $\nu^c\nu^c$-type Higgs field and its conjugate are used, the combined tube in \Fig{dumbbell_afterEW_422} splits into two equal ones as shown in \Fig{splittube422}. We can connect the loose ends in \Fig{splittube422} to construct necklaces of all kinds which can contain both red and blue monopoles and their antimonopoles. Each red monopole can connect with its antimonopole or a blue monopole, and similarly each blue monopole with its antimonopole or a red one. An example is displayed in \Fig{necklace422}, where the brown monopole is the antimonopole of the blue monopole and the green monopole is the antimonopole of the red one. The overall Coulomb magnetic flux of a necklace is a multiple of the Schwinger monopole flux. In particular, the flux of the necklace in \Fig{necklace422} is minus the Schwinger flux. One can imagine a variety of such necklaces which can all be considered as realizations of the $\Z_2$ string in the following sense. If one inverts the arrows in all their lines, the sign of their overall Coulomb flux changes. However, by adding an appropriate number of Schwinger monopoles on some of their beads, we can make them the same as before. Thus, we consider these necklaces as $\Z_2$ strings but with Schwinger monopoles attached to them.

A subsequent breaking of the $\Z_2$ subgroup of $\Z_4$, the center of $SO(10)$, by the VEV of a $\nu^c$-type Higgs field and its conjugate causes the necklaces to become boundaries of $\Z_2$ domain walls (see section \ref{subsec:Wall_bounded_Necklaces}).

\section{$\mathbf{SO(10)}$ breaking via Flipped $\mathbf{SU(5)}$}
\label{sec:flipSU5}
We now turn to the discussion of the breaking of $SO(10)$ via its flipped subgroup $SU(5)\times U(1)_X$ \cite{Barr:1981qv}. Consider the following breaking chain:
\begin{equation}
\begin{split}
     SO(10)&\longrightarrow SU(5)\times U(1)_X\\
     &\longrightarrow SU(3)_c\times SU(2)_L \times U(1)_Y\\
     &\longrightarrow SU(3)_c \times U(1)_{em}. 
\end{split}
\label{flipchain}
\end{equation}
Under the flipped subgroup $SU(5) \times U(1)_X$ of $SO(10)$, the identification of the fermion states in the $16$-plet is as follows:
\begin{equation}
    16=1(-5)+\overline{5}(3)+10(-1)
\label{16flip}
\end{equation}
with
\begin{equation}
    \begin{split}
    &\mspace{-14mu}\underbrace{1}_{e^c}\\
    &\overline{5}=\underbrace{(1,2)(-3)}_{\begin{pmatrix} \nu \\ e \end{pmatrix} }+\underbrace{(\overline{3},1)(2)}_{u^c} \\
    &10=\underbrace{(1,1)(6)}_{\nu^c}+\underbrace{(\overline{3},1)(-4)}_{d^c}+\underbrace{(3,2)(1)}_{\begin{pmatrix} u \\ d\end{pmatrix} }
    \end{split}
\label{5bar10_321flip}
\end{equation}
under $SU(3)_c\times SU(2)_L \times U(1)_Z \subset SU(5)$.

\begin{figure}[h]
\centering

      \includegraphics[width=0.3\textwidth,angle=0]{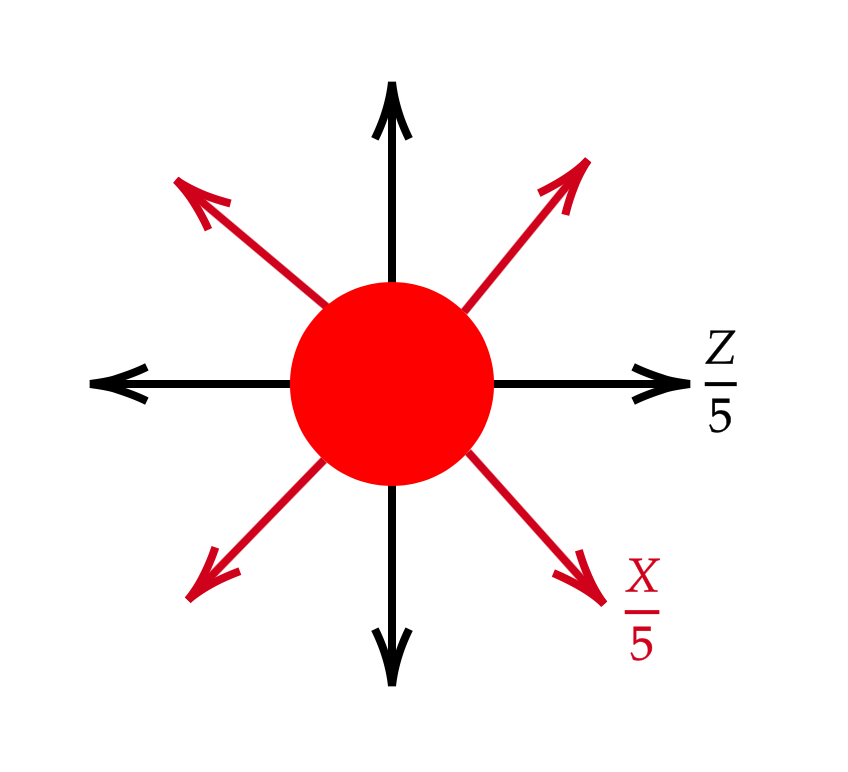}
        \caption{The Coulomb magnetic fluxes corresponding to $2\pi/5$ rotations around $U(1)_X$ and $U(1)_Z$ carried by the monopole generated from the breaking of $SO(10)$ to $SU(5)\times U(1)_X$ by the VEV of an $SO(10)$ $45$-plet.}
  \label{UX_UZ_monopole_flipped}
\end{figure}

The first breaking in Eq.~(\ref{flipchain}) is achieved by the VEV of an $SO(10)$ $45$-plet and produces a monopole which carries Coulomb magnetic fluxes corresponding to $2\pi/5$ rotations around $U(1)_X$ and $U(1)_Z$ as indicated in \Fig{UX_UZ_monopole_flipped}.

\subsection{Dumbbell configuration}
The breaking of $U(1)_X \times U(1)_Z$ is achieved by the VEV of a $\nu^c$-type Higgs field contained in the $SU(5)$ $10$-plet and its conjugate. The unbroken combination is $Z+6X$. To find the orthogonal broken generator, we define the GUT normalized generators 
\begin{equation}
    Q_Z=\frac{1}{\sqrt{60}}Z,\;\;  Q_X=\frac{1}{2\sqrt{10}}X,
\label{QZQX}
\end{equation}
such that 
\begin{equation}
    Z+6 X=\sqrt{60} (Q_Z+2\sqrt{6}Q_X)
\label{Z+6X}
\end{equation}
and the broken generator is thus proportional to $-4Z+X$.
It is easy to check that all the $Z+6X$ and $-4Z+X$ charges in the 16-plet are proportional to 5 and that by dividing these generators by 5, we obtain the smallest possible integer charges and thus periodicity $2\pi$ of the corresponding $U(1)$'s. Thus, we define the unbroken $(Y)$ and broken $(\Tilde{Y})$ generators as 
\begin{equation}
    Y=\frac{Z+6X}{5}, \;\; \Tilde{Y}=\frac{-4Z+X}{5}.
\label{YandYtilde}
\end{equation}
 To see if there are common elements between $U(1)_Y$ and $U(1)_{\Tilde{Y}}$, we solve the equation 
 \begin{equation}
     \alpha \Tilde{Y}=\beta Y (mod\; 2\pi)\;\;\;\text{on all states in the 16-plet}
\label{commonYandYtilde}
 \end{equation}
and obtain two independent equations 
\begin{equation}
    5\alpha=0\;(mod\;2\pi),\;\;\; \alpha=\beta\;(mod\;2\pi).
\label{solncommonYandYtilde}
\end{equation}
It is clear that the $\Z_5$ subgroups of $U(1)_Y$ and $U(1)_{\Tilde{Y}}$ coincide. Moreover,
\begin{equation}
    \frac{2\pi}{5}X=-\frac{2\pi}{5}Z+\frac{2\pi}{5}(Y-\Tilde{Y})
\label{Z5coincide}
\end{equation}
and $Y-\Tilde{Y}$ is a multiple of 5 on all states in 16. Thus,
\begin{equation}
    \frac{2\pi}{5}X=-\frac{2\pi}{5}Z\;(mod\;2\pi)\;\;\;\text{on all states}
\label{YtildeZrelat}
\end{equation}
and 
\begin{equation}
    e^{i\frac{2\pi}{5}\Tilde{Y}}=e^{i\frac{2\pi}{5}Y}\;\;\;{\rm and}\;\;\; e^{i\frac{2\pi}{5}X}=e^{-i\frac{2\pi}{5}Z},
\label{YtildeYXZrelat}
\end{equation}
i.e. the $\Z_5$ subgroups of $U(1)_{\Tilde{Y}}$, $U(1)_Y$ coincide, and the $\Z_5$ subgroup of $U(1)_X$ coincides with the $\Z_5$ subgroup of $U(1)_Z$ (=center of $SU(5)$), but with inverted generator. The generator $e^{i\pi Y}$ of the $\Z_2$ subgroup of $U(1)_Y$ acts as minus identity on all the $SU(2)_L$ doublets in 16, and as an identity element on all $SU(2)_L$ singlets. In other words, it coincides with the generator of the $\Z_2$ center of $SU(2)_L$. Also $e^{iY{2\pi}/{3}}$ acts like $e^{\mp i {2\pi}/{3}}\;1$ on the $SU(3)_c$ triplets (antitriplets), and as an identity element on $SU(3)_c$ singlets, and so it coincides with the inverse generator of the center of $SU(3)_c$. In conclusion, the $\Z_6=\Z_3\times \Z_2$ subgroup of $U(1)_Y$ lies in $SU(3)_c \times SU(2)_L$.

\begin{figure}[h]
\centering

      \includegraphics[width=0.55\textwidth,angle=0]{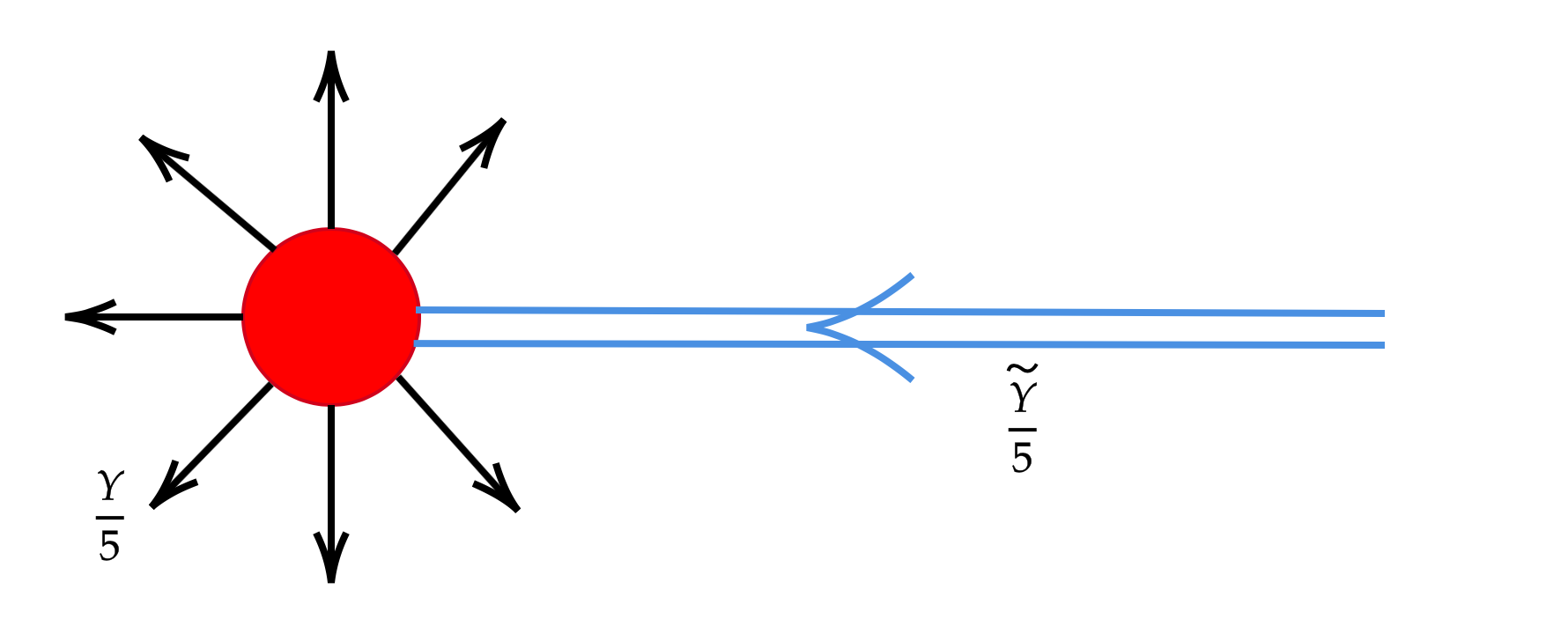}
        \caption{A $U(1)_{\Tilde{Y}}\; \Z_5$ magnetic tube emerges from the monopole in \Fig{UX_UZ_monopole_flipped} as a result of the breaking of $\Tilde{Y}$ in the second step of Eq.~(\ref{flipchain}).}
  \label{coulombtubeflip}
\end{figure}

\begin{figure}[h]
\centering

      \includegraphics[width=0.55\textwidth,angle=0]{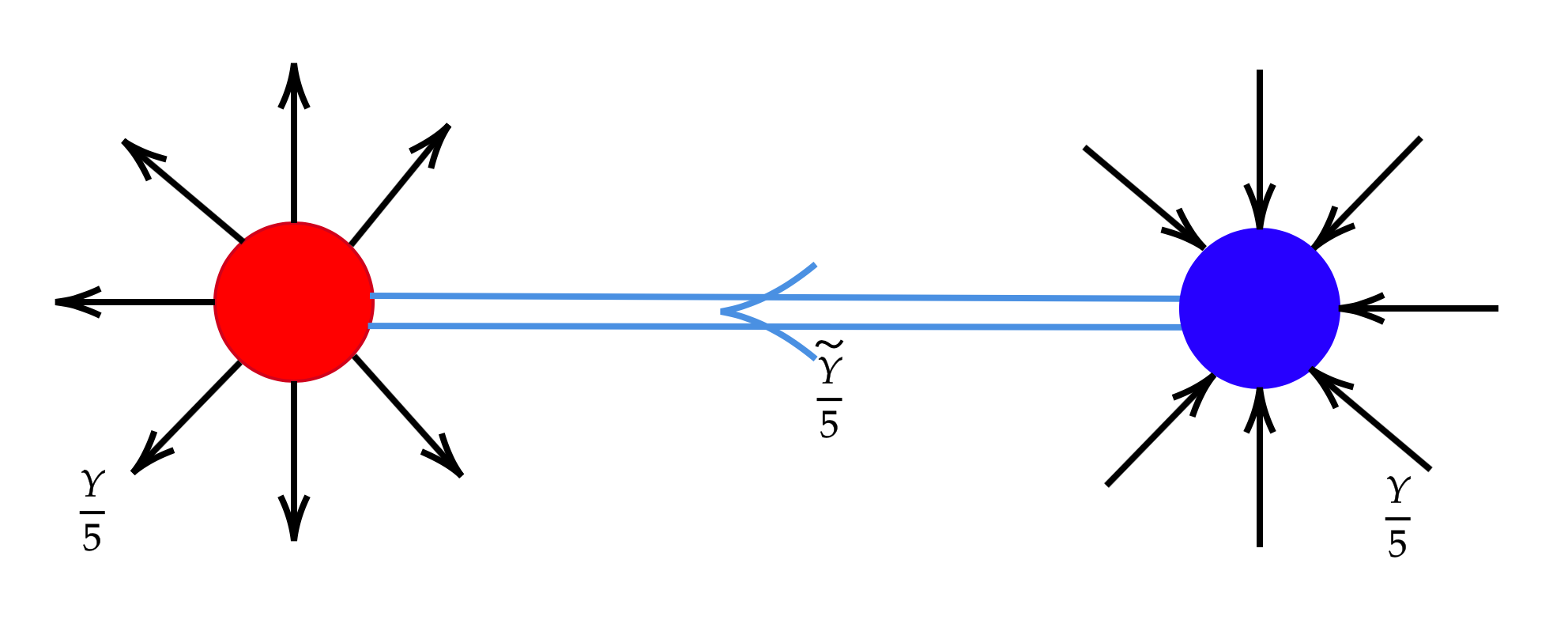}
        \caption{Monopole (red)-antimonopole (blue) pair bound together via a flux tube.}
  \label{dumbbellflip}
\end{figure}

From Eq.~(\ref{YandYtilde}) we find that
\begin{equation}
X=\frac{4Y+\Tilde{Y}}{5},\;\;\;Z=\frac{Y-6\Tilde{Y}}{5}
\label{XZintermsYYtilde}
\end{equation}
and the magnetic flux of the monopole in \Fig{UX_UZ_monopole_flipped} is 
\begin{equation}
    \frac{1}{5}(X+Z)=\frac{1}{5}(Y-\Tilde{Y}).
\label{fluxX+Z}
\end{equation}
At the second step of symmetry breaking, $\Tilde{Y}$ is broken and a $U(1)_{\Tilde{Y}}\; \Z_5$ tube appears (see \Fig{coulombtubeflip}).
Going around the tube in the positive direction, the VEV of the $\nu^c$-type Higgs field changes phase by $-2\pi$. A monopole and an antimonopole can be bound together via a tube to form a dumbbell as depicted in \Fig{dumbbellflip}.

The EW Higgs doublets belong to an $SO(10)$ $10$-plet and a $\overline{126}$-plet with the following identification: Under $SU(5)\times U(1)_X$,
\begin{equation}
    \begin{split}
     10 &= 5(2)+\overline{5}(-2) \\
    \overline{126} &= 5(2)+\overline{45}(-2)+...
    \end{split}
\label{10SU5flip}
\end{equation}
with
\begin{equation}
    \begin{split}
    5&=\underbrace{(1,2)(3)}_{h_d}+...,\\
    \overline{5}&=\underbrace{(1,2)(-3)}_{h_u}+...,\\
    \overline{45}&=\underbrace{(1,2)(-3)}_{h_u}+...
    \end{split}
\label{55barSU5_321_flip}
\end{equation}
under $SU(3)_c\times SU(2)_L\times U(1)_Z$.
\begin{figure}[h]
\centering

      \includegraphics[width=0.48\textwidth,angle=0]{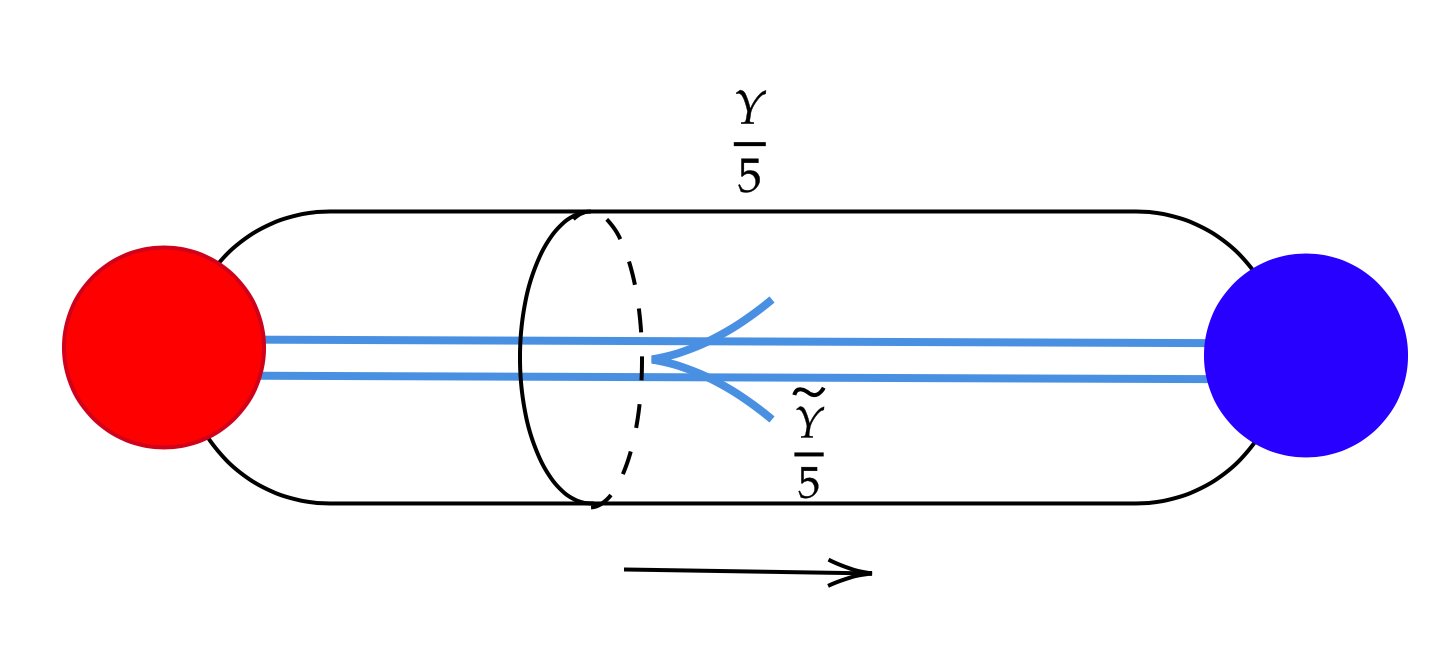}
        \caption{Formation of an EW flux tube by the $Y/5$ flux in the dumbbell in \Fig{dumbbellflip}.}
  \label{dumbbell_afterEW_flip}
\end{figure}

Going around the $\Tilde{Y}/5$ tube in \Fig{dumbbellflip} in the positive direction the phase of the VEV of $h_d$ $(h_u)$ change by $−4\pi/5$ $(+6\pi/5)$. The Y/5 Coulomb flux in the dumbbell will then form an EW tube as shown in \Fig{dumbbell_afterEW_flip}. Going around this combined tube in the positive direction with respect to the $\Tilde{Y}/5$ tube, the phases of these VEVs pick up an extra contribution of $-6\pi/5$ $(+4\pi/5)$ and thus their total change is $-2\pi$ $(+2\pi)$. The phases of the VEVs of the $\nu^c$-, $\overline{\nu^c}$-type Higgs fields, on the other hand, remain unaffected by the EW flux tube and still change by $-2\pi$, $2\pi$. 

The second breaking in Eq.~(\ref{flipchain}) does not produce monopoles since the second homotopy group of the vacuum manifold
\begin{equation}
    \pi_2\left(\frac{SU(5)\times U(1)_X/\Z_5}{SU(3)_c \times SU(2)_L \times U(1)_Y / \Z_6}  \right) = \{1\}.
\label{homotopypi2_flip}
\end{equation}

\subsection{Unconfined $SO(10)$ monopole}
We will now see how the unconfined SO(10) monopole required for charge quantization comes about. To this end, we write
\begin{equation}
    e^{-\frac{2\pi i}{6}Z}=e^{-i \pi Z}\;e^{\frac{2\pi i}{3}Z}.
\label{expZ_flip}
\end{equation}
The element $e^{-i\pi Z}$ acts on the doublet in $5=(1,2)(3)+(3,1)(-2)$ as minus identity, but leaves invariant the triplet. Consequently, $e^{-i \pi Z}=e^{i \pi T_L^3}$, the generator of the $\Z_2$ center of $SU(2)_L$. The element $e^{2\pi iZ/3}$ acts as an identity element on the doublet, and on the triplet as $e^{2\pi iT^8_c/3}$, which is the generator of the $\Z_3$ center of $SU(3)_c$. Thus,
\begin{equation}
    e^{-\frac{2 \pi i}{6}Z}=e^{i \pi T_L^3}\;e^{\frac{2 \pi i}{3}T_c^8}.
\label{expT_flip}
\end{equation}
Therefore, there exists a loop in $SU(5)$ which starts from unity and moves along $U(1)_Z$ in the negative direction until $−2\pi/6$, and then continues from the generators of the $\Z_2$ and  $\Z_3$ centers of $SU(2)_L$ and $SU(3)_c$ back to unity. This is a contractable loop (as all loops) in $SU(5)$. It is thus legitimate, for unbroken $SU(5)$, to add the corresponding flux to the monopole in \Fig{UX_UZ_monopole_flipped}.
The flux of the monopole then becomes
\begin{equation}
\begin{split}
     \frac{Z}{5}+\frac{X}{5}-\frac{Z}{6}+\;\text{$SU(2)_L$ flux corresponding to the generator of $\Z_2$}\\
    +\;\text{$SU(3)_c$ flux corresponding to the inverse generator of $\Z_3$}.
\end{split}
\label{fluxwithSU2_flip}
\end{equation}
\begin{figure}[h!]
\centering

      \includegraphics[width=0.4\textwidth,angle=0]{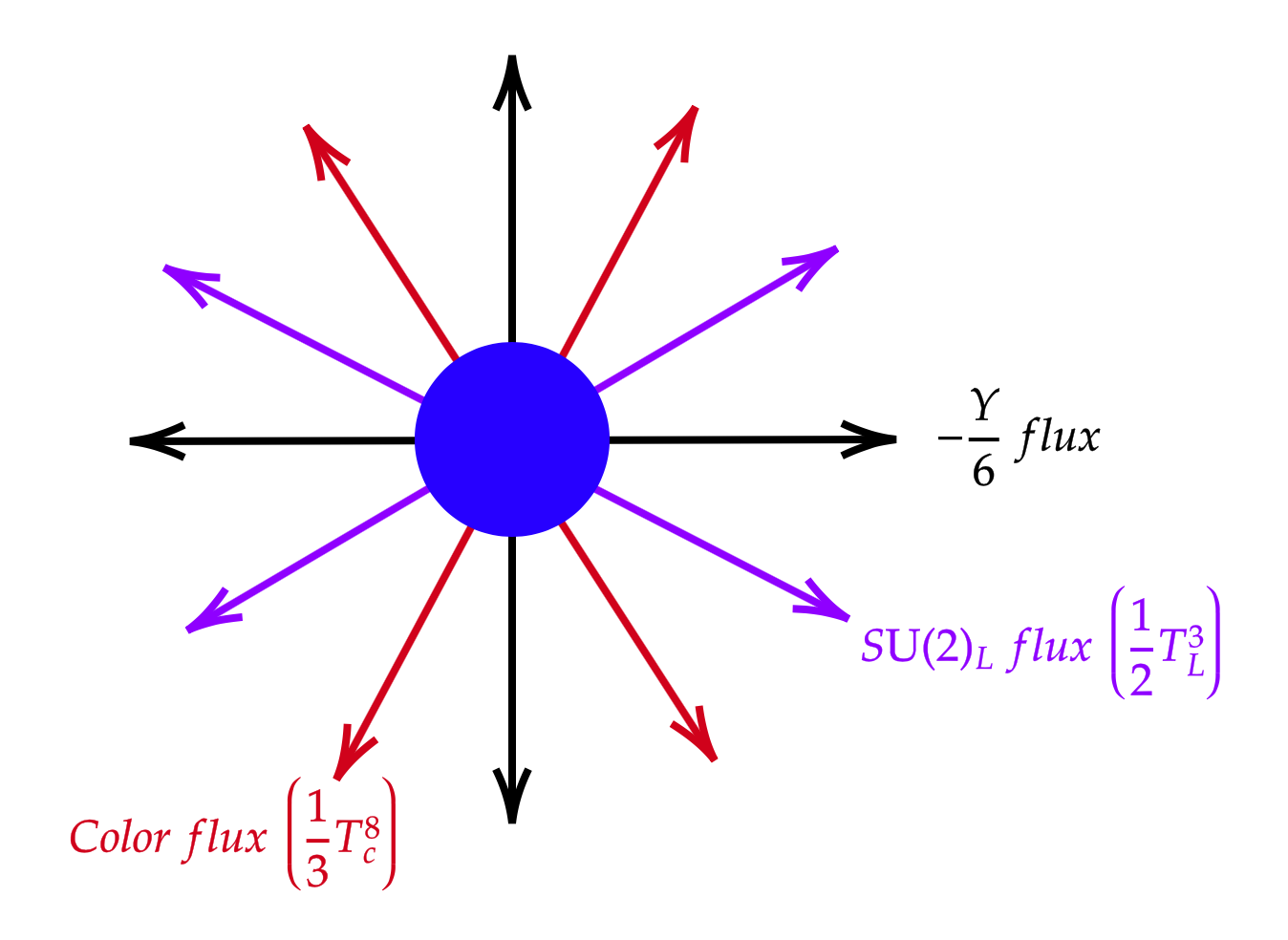}
        \caption{The abelian Coulomb flux of the antimonopole, along with color and $SU(2)_L$ fluxes, prior to EW breaking.}
  \label{monopole_SU2flux_flip}
\end{figure}
However, $Z/5+X/5-Z/6=Y/6$, which is unbroken by the VEVs of the $\nu^c$- and $\overline{\nu^c}$-type Higgs fields and remains Coulombic. Therefore, prior to the EW breaking, the monopole carries Coulomb flux $Y/6$ plus color and $SU(2)_L$ fluxes. This is shown in \Fig{monopole_SU2flux_flip} after inverting all the fluxes for later convenience. After the EW breaking the electric charge operator
\begin{equation}
    Q=-\frac{1}{6} Y + \frac{1}{2} T_L^3
\label{QafterEW_flip}
\end{equation}
remains unbroken as it should and the antimonopole takes the form shown in \Fig{monopole_colorflux_afterEW_flip}, which is the unconfined $SO(10)$ monopole required for the quantization of electric charge. The unconfined $SO(10)$ monopole can be also constructed if we break $SU(5)$ by the VEV of a $24$-plet in an $SO(10)$ $45$-plet prior to the second breaking step in Eq.~(\ref{flipchain}). In this case, we obtain a monopole as in \Fig{unconfinedSO10_flip}. Note that $Z/6 =Y/30-\Tilde{Y}/5$, which implies that, after the VEVs of the $\nu^c$-, $\overline{\nu^c}$-type Higgs fields, this monopole takes the form shown in \Fig{monopolecolorflux_tube_flip}.

\begin{figure}[h]
\centering

      \includegraphics[width=0.35\textwidth,angle=0]{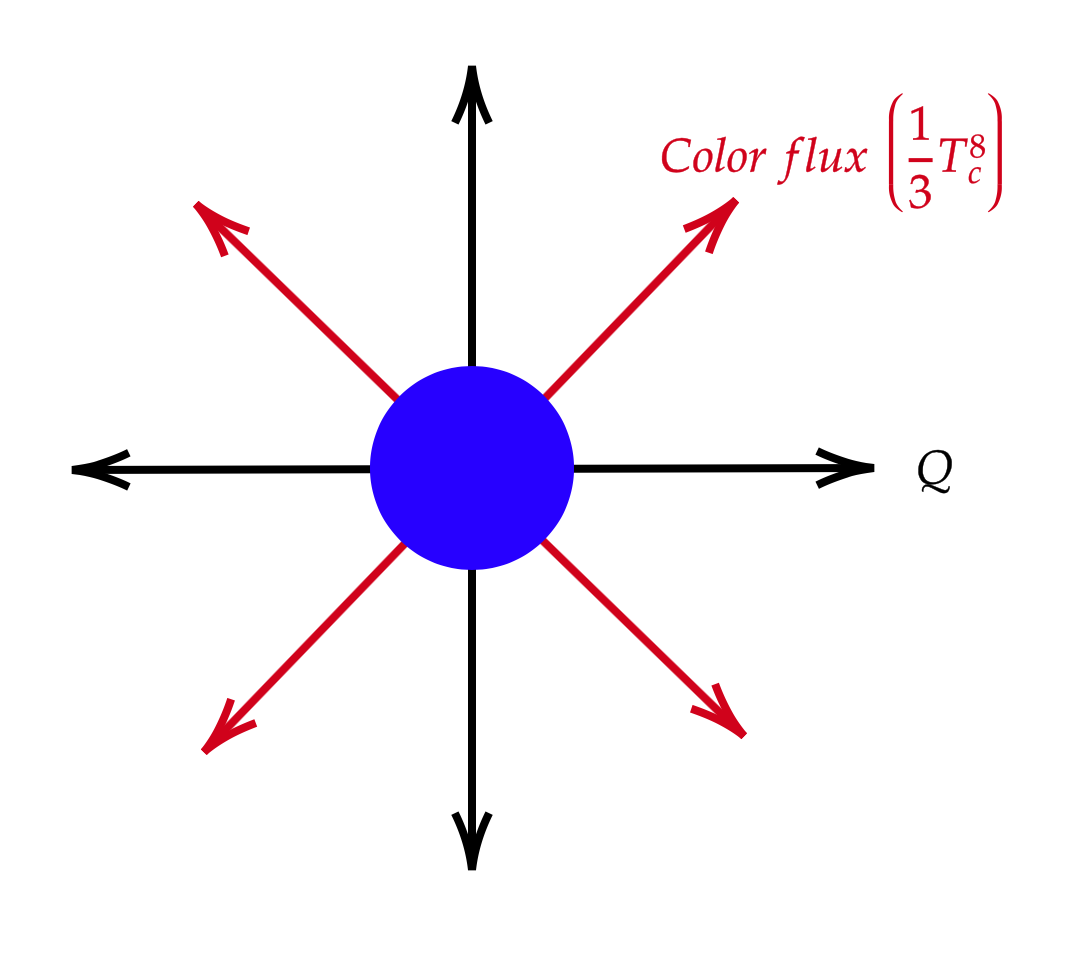}
        \caption{The antimonopole in \Fig{monopole_SU2flux_flip} after EW breaking. This is the unconfined $SO(10)$ monopole.}
  \label{monopole_colorflux_afterEW_flip}
\end{figure}
\begin{figure}[h]
\centering

      \includegraphics[width=0.35\textwidth,angle=0]{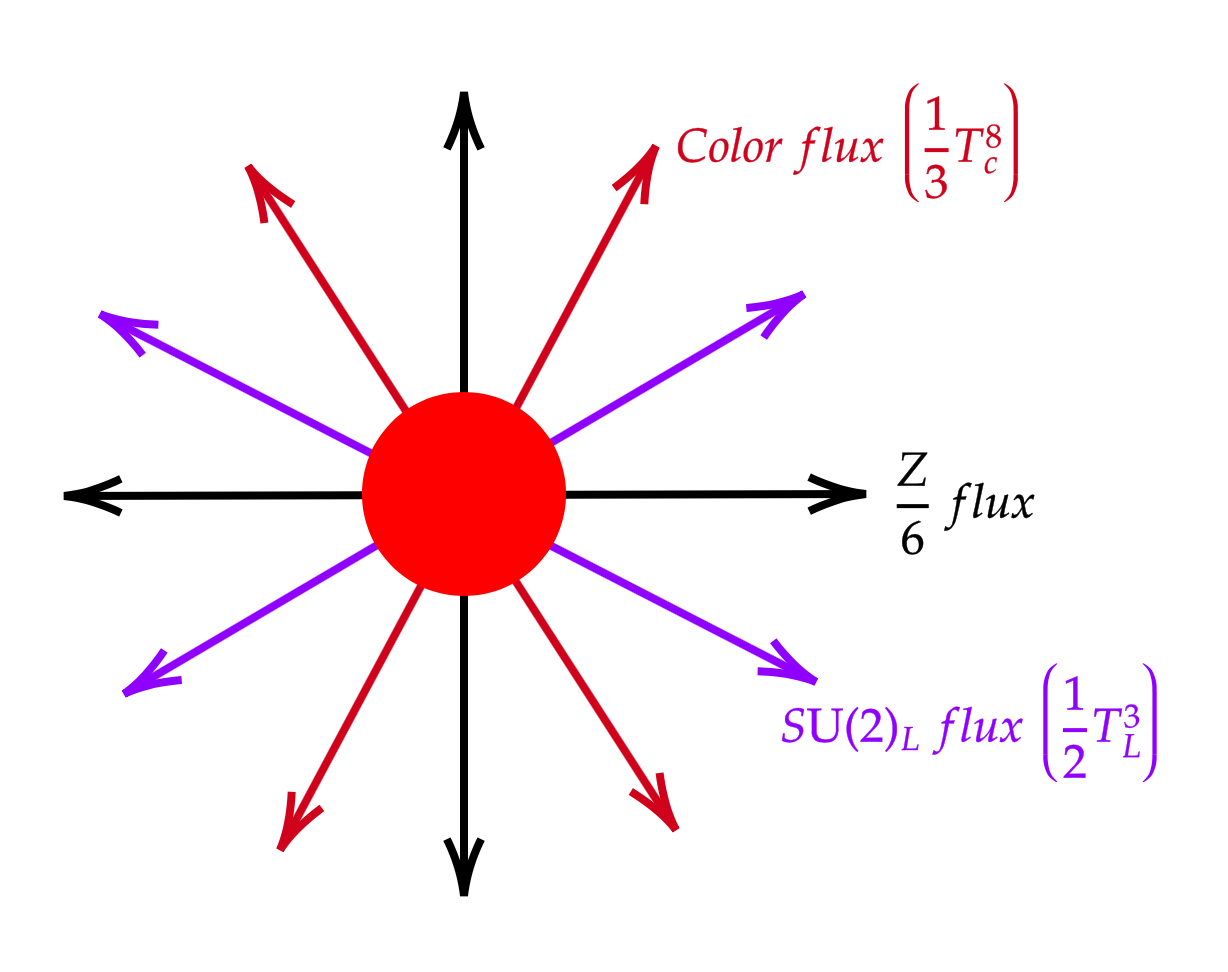}
        \caption{Monopole generated by breaking $SU(5)$ via the VEV of a $24$-plet in an $SO(10)$ $45$-plet, prior to the second breaking step of Eq.~(\ref{flipchain}).}
  \label{unconfinedSO10_flip}
\end{figure}
\begin{figure}[h]
\centering

      \includegraphics[width=0.6\textwidth,angle=0]{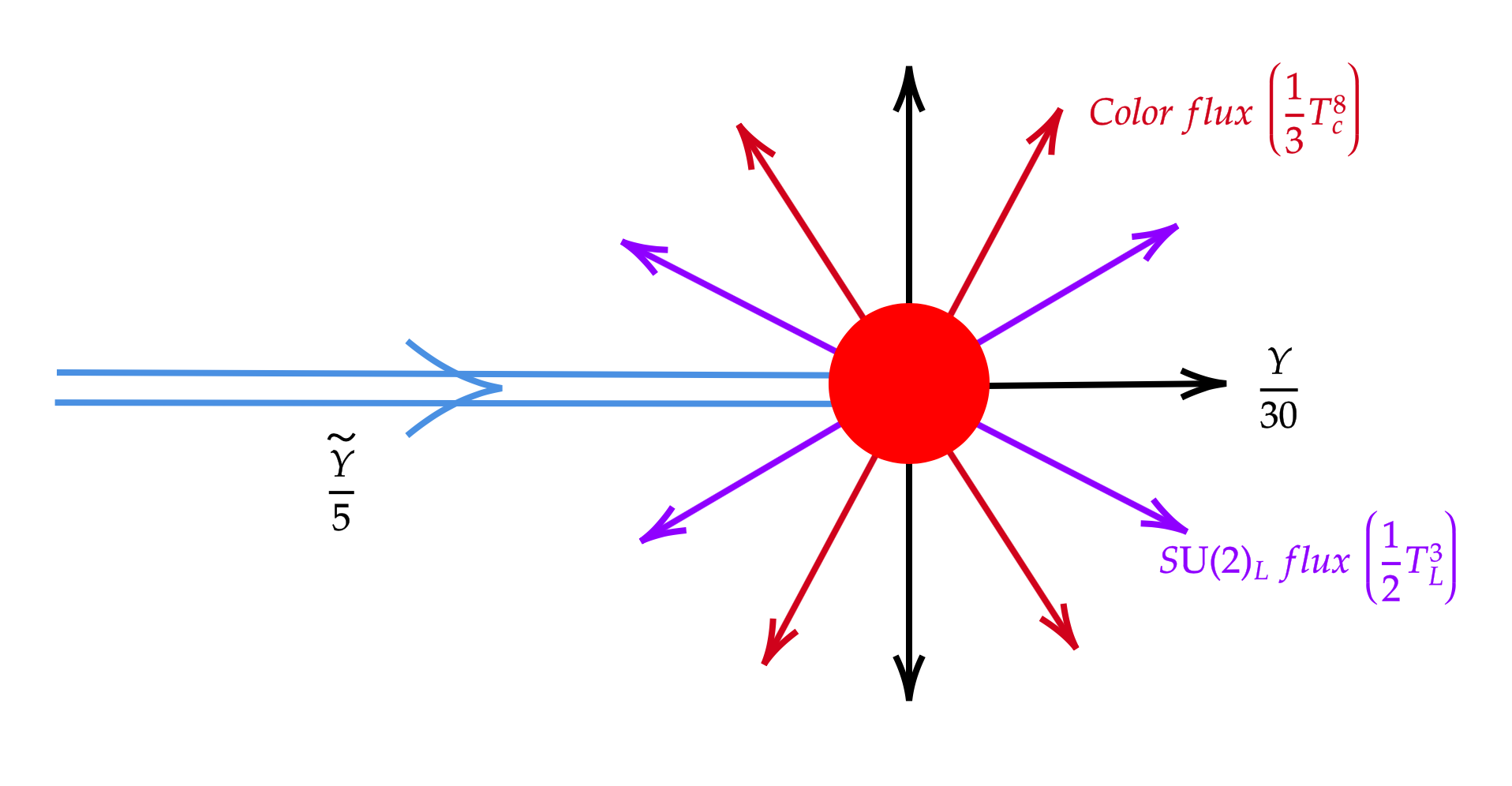}
        \caption{The monopole in \Fig{unconfinedSO10_flip} after the VEVs of the $\nu^c$-, $\overline{\nu^c}$-type Higgs fields.}
  \label{monopolecolorflux_tube_flip}
\end{figure}
\begin{figure}[h]
\centering

      \includegraphics[width=0.65\textwidth,angle=0]{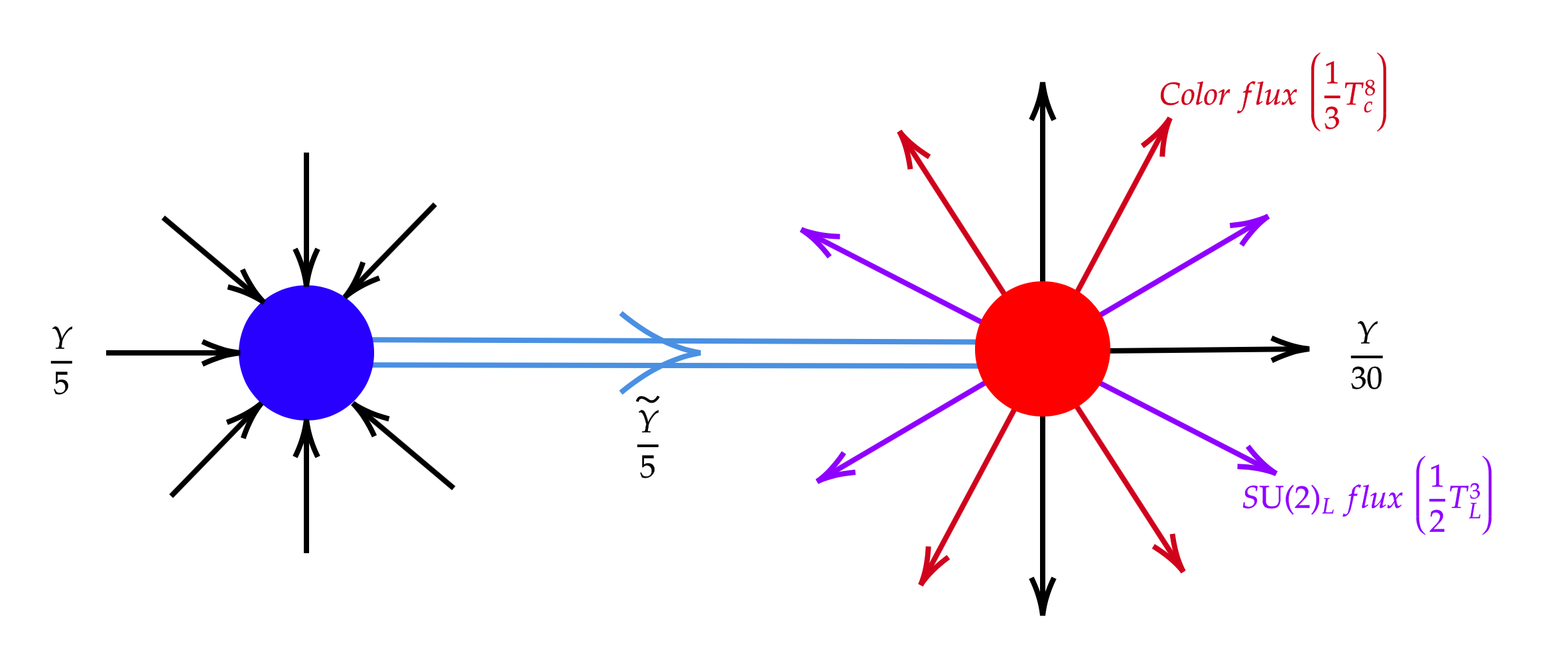}
        \caption{Dumbbell configuration obtained by combining the monopole of \Fig{monopolecolorflux_tube_flip} with the antimonopole analogue of \Fig{coulombtubeflip}.}
  \label{dumbbell_colorflux_flip}
\end{figure}
Combining this monopole with the antimonopole analogue of \Fig{coulombtubeflip}, we obtain the dumbbell in \Fig{dumbbell_colorflux_flip} which, after the merger of the two, reduces to the monopole in \Fig{monopole_SU2flux_flip}, and after the EW breaking to the unconfined $SO(10)$ monopole in \Fig{monopole_colorflux_afterEW_flip}. Breaking the EW symmetry before the contraction of the $\Tilde{Y}/5$ tube, the $Y/5$ flux of the antimonopole (blue) in \Fig{dumbbell_colorflux_flip} forms an EW tube like in \Fig{dumbbell_afterEW_flip}, and the second monopole becomes like the one in \Fig{monopole_colorflux_afterEW_flip}.

\subsection{Necklace configuration}
\begin{figure}[h!]
\centering

      \includegraphics[width=0.45\textwidth,angle=0]{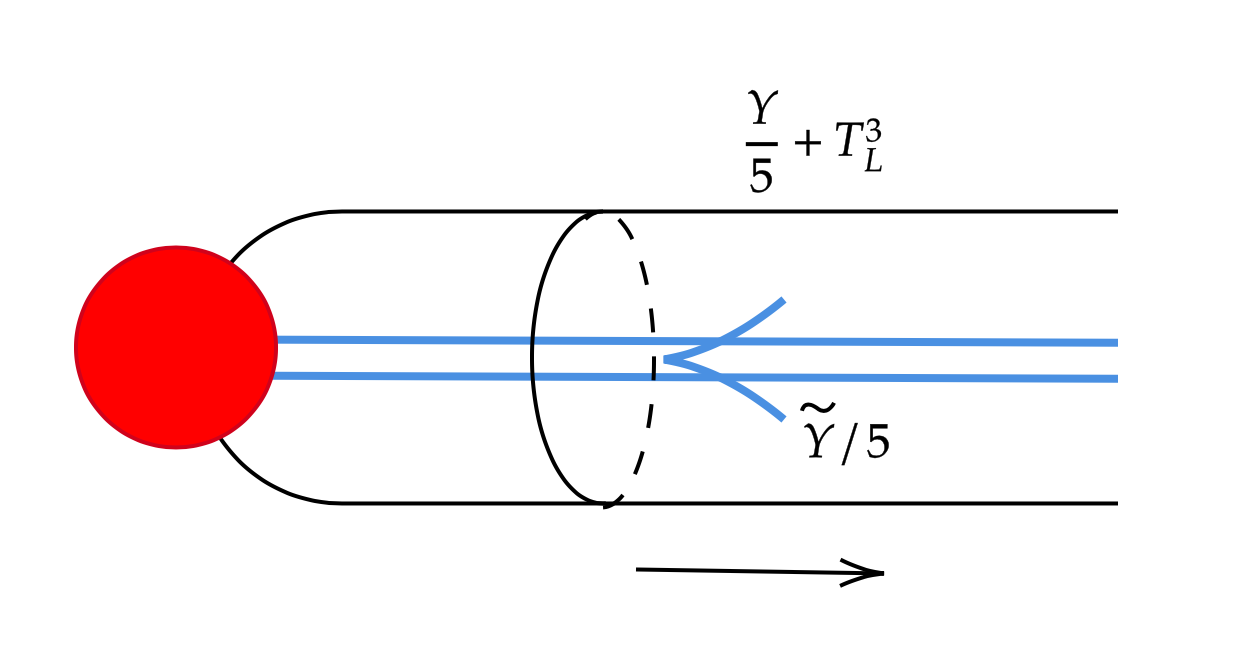}
        \caption{The $SO(10)$ monopole in \Fig{coulombtubeflip} after the EW breaking with an extra full $T^3_L$ flux.}
  \label{coulombtube_afterEW_flip}
\end{figure}

\begin{figure}[h!]
\centering

      \includegraphics[width=0.55\textwidth,angle=0]{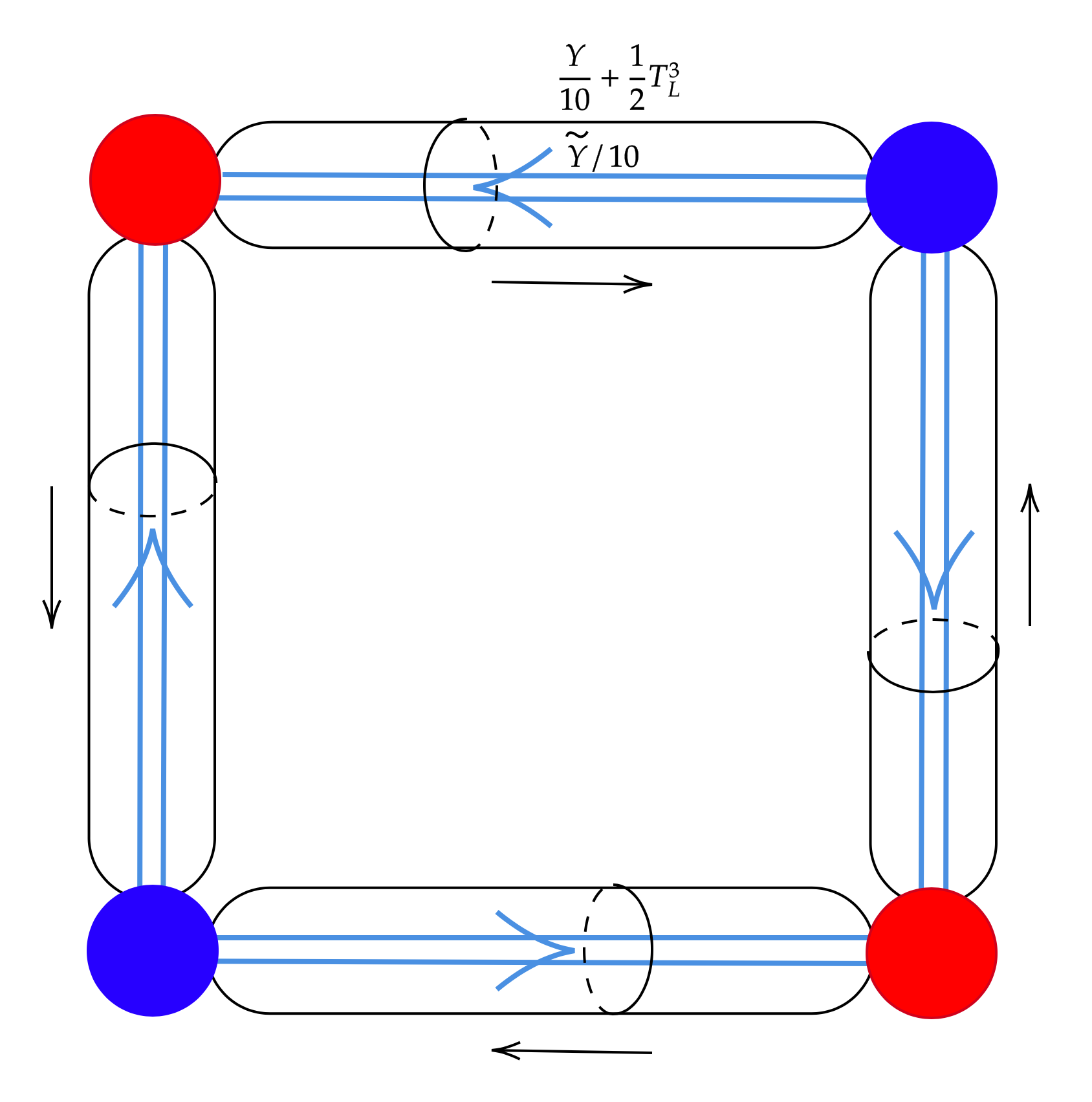}
        \caption{The tube shown in \Fig{coulombtube_afterEW_flip} can be split into two tubes and consequently blue and red monopoles can be connected through these tubes to form a necklace, which is a realization of $\Z_2$ string.}
  \label{necklaceflip}
\end{figure}

We will now discuss the case with topologically stable $\Z_2$ strings in flipped SU(5), with the breaking of $SU(5) \times U(1)_X$ achieved by the VEV of the $\nu^c\nu^c$-type Higgs field in the $126$-plet of $SO(10)$ (and its conjugate) identified as follows:
\begin{equation}
    \begin{split}
        126=&\overline{50}(-2)+...\;\;\;\text{under $SU(5)\times U(1)_X$},\\
        &\overline{50}=\underbrace{(1,1)(12)}_{\nu^c\nu^c}+...\;\;\;\text{under $SU(3)_c\times SU(2)_L \times U(1)_Z$}.
    \end{split}
\label{126_flip_50bar_321}
\end{equation}
This leaves the $\Z_2$ subgroup of the center of $SO(10)$ unbroken and leads to $\Z_2$ strings, in agreement with the result proved in Ref.~\cite{Kibble:1982ae} (see also Refs.~\cite{Hindmarsh:1985xc,Everett:1986eh,Jeannerot:2003qv,Dror:2019syi}). Consider the monopole in \Fig{coulombtubeflip} and add a full $T^3_L$ Coulomb flux emerging from it. After the EW breaking this monopole takes the form shown in \Fig{coulombtube_afterEW_flip}.
Note that $Y/5 + T^3_L$ is orthogonal to $Q = −Y/6 + T^3_L/2$, as one can show using the normalized generators. Going around this tube in the positive direction with respect to the $\Tilde{Y}/5$ tube, we find that the phase of the VEV of $h_d$ $(h_u)$ acquires an additional contribution of $−2\pi$ $(2\pi)$ from $T^3_L$. Therefore the total change is $−4\pi$ $(4\pi)$. The phase of $\nu^c\nu^c$-type Higgs field, of course, changes by $−4\pi$. This allows us to split the tube into two equal tubes and construct necklaces like the one shown in \Fig{necklaceflip}, which correspond to $\Z_2$ strings in the flipped $SU(5)$ model. Note that their construction requires the use of the confined version of the monopole.

A subsequent breaking of the $\Z_2$ subgroup of $\Z_4$, the center of $SO(10)$, by the VEV of a $\nu^c$-type Higgs field and its conjugate causes the necklaces to become boundaries of $\Z_2$ domain walls (see section \ref{subsec:Wall_bounded_Necklaces}).

\section{Conclusion}
In this paper we have focused on a variety of (new) composite topological structures whose presence in the early universe may lead to a set of distinct signatures, which can be experimentally tested in the ongoing and future observations. Such observations would provide a novel way to probe the symmetry breaking pattern of the underlying GUT, and indeed a unique way if the scales involved are of magnitude well beyond the reach of any future collider. The observability of the stochastic gravitational wave background generated by our composite objects depends on various factors. We suggest the future study of the $SU(4)_c\times SU(2)_L\times SU(2)_R$ case with $\Z_2$ strings, under the assumption that the red/blue monopoles are partially inflated and re-enter the horizon after the equidensity time $t_{eq}$. The strings and, consequently, the loops formed during radiation dominance are then expected to have smaller tension than the necklaces since they will be mostly without monopoles on them. The contribution of the loops created before $t_{eq}$ and decaying after it generates the low-frequency peak in the gravity wave spectrum \cite{Chakrabortty:2020otp}. Therefore, the fact that they have smaller tension may help with satisfying the PTA bound. Loops created after $t_{eq}$ will have larger tension due to the monopole re-entrance in the horizon but do not contribute significantly to the main low-frequency peak. Finally, it would be exciting indeed if some of these composite structures could be found in condensed matter systems.

\acknowledgments

A.T. is partially supported by the Bartol Research Institute, University of Delaware. The work of G.L. and Q.S. is supported by the Hellenic Foundation for Research and Innovation (H.F.R.I.) under the “First Call for H.F.R.I. Research Projects to support Faculty Members and Researchers and the procurement of high-cost research equipment grant” (Project Number:2251). Q.S. thanks Rinku Maji and Anish Ghoshal for useful discussions.

\bibliographystyle{JHEP}

\end{document}